\documentclass[reprint,amsmath,amssymb,aps]{revtex4-2}

\usepackage{array}
\usepackage{graphicx}
\usepackage{dcolumn}
\usepackage{bm}

\DeclareMathOperator{\Tr}{Tr}

\begin{document}

\title{Diffusio-phoretic fast swelling of chemically responsive hydrogels}

\author{Chinmay Katke$^{1, 2\,}$, Peter A. Korevaar$^{3\,}$, and C. Nadir Kaplan$^{1, 2}$}

\email{nadirkaplan@vt.edu}
\affiliation{$^{1}$Department of Physics,
 Virginia Polytechnic Institute and State University,
 Blacksburg, VA 24061, USA,\\
 $^{2}$Center for Soft Matter and Biological Physics, Virginia Polytechnic Institute and State University,
Blacksburg, VA 24061, USA,\\
$^{3}$Institute for Molecules and Materials,
Radboud University, 6525 AJ Nijmegen, The Netherlands.}

%TC:ignore

\begin{abstract}
Acid-induced release of stored ions from polyacrylic acid hydrogels (with a free surface fully permeable to the ion and acid flux) was observed to increase the gel osmotic pressure that leads to rapid, temporary swelling faster than the characteristic solvent absorption rate of the gel. Here we develop a continuum poroelastic theory that quantitatively explains the experiments by introducing a ``gel diffusio-phoresis" mechanism: Steric repulsion between the gel polymers and released ions can induce a diffusio-osmotic solvent intake counteracted by the diffusio-phoretic expansion of the gel network. For applications ranging from drug delivery to soft robotics, engineering the gel diffusio-phoresis may enable stimuli-responsive hydrogels with amplified strain rates and power output.
\end{abstract}

%TC:endignore

\maketitle

The capability of osmosis to convert modest concentration differences into significant pressures underlies biological processes ranging from turgor pressure regulation in walled cells to the urea-water separation in the kidney, which inspired many applications for chemo-mechanical energy conversion~\cite{Kanahama2023, Schulgasser1997,GOTTSCHALK1964670,STEPHENSON197285,Brogioli2009,TUFA2018290, GEISE2011130,Werber2016,Logan2012,Siria2013,Siria2017,Kedem1961}. It also governs hydrogel expansion through solvent imbibition, and has enabled stimuli-responsive micron-scale gels for, e.g., soft actuation or synthetic homeostasis~\cite{Zhang2017,Ulijn2007,Shim2012,Khiabani2022,Yuk2017,Ma2013,Dong2006, Zabow2015,Skoge2014,Nerbonne2005,Peter2017}. Scaling up these gel designs, however, is hindered by the drastic reduction of their strain rate and power output since a gel with the shortest dimension $H$ and permeability $k_f$ typically absorbs the solvent at a rate $\tau^{-1}\sim k_f/H^2\,.$ Although increasing the pore size (i.e., higher $k_f$) mitigates this limitation, it reduces the density of the gel polymer network, compromising on its functionalization and in turn the gel responsiveness to external fields for effective chemo-mechanical energy transduction~\cite{Choudhary2022,epstein2014,Arens2017}.    

\begin{figure}[t!]
\includegraphics[width=1\columnwidth]{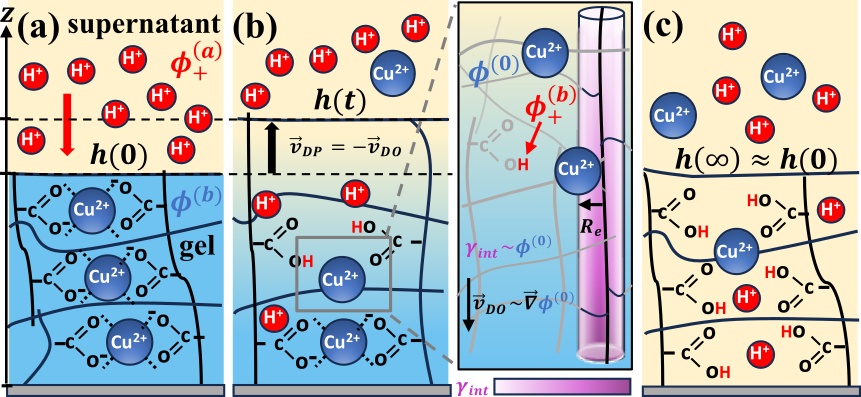} 
\caption{\textbf{PAA gel response to competing stimuli.} \textbf{(a)} Acid (red, volume fraction $\phi^{(a)}_+$) is delivered from the supernatant solution into a copper-laden PAA hydrogel (attached to a substrate) with a contracted initial height $h(0)<H$ due to the chelation between COO$^-$ and Cu$^{2+}$ (blue, volume fraction $\phi^{(b)}$), which turns the gel blue. \textbf{(b)} The formation of COOH groups (volume fraction $\phi^{(b)}_+$) releases Cu$^{2+}$ with a volume fraction $\phi^{(0)}$ into the gel solution. The gel swells with a time-dependent height $h(t)>h(0)$ and becomes colorless while a gradient $\nabla\phi^{(0)}$ along the $-z$ axis emerges~\cite{Korevaar2020}. The diffusio-phoretic swelling velocity $\mathbf{v}_{DP}$ negates the diffusio-osmotic solvent velocity $\mathbf{v}_{DO}$ whose origin is shown in the inset:  The surface tension $\gamma_{int}$ due to the steric repulsion between the gel backbone and copper ions at a core distance $R_e$ is proportional to $\phi^{(0)}\,,$ whose gradient sets the diffusio-osmotic flow direction. \textbf{(c)} The copper gradient and $\mathbf{v}_{DO}\,, \mathbf{v}_{DP}$ eventually vanish due to Cu$^{2+}$ diffusion, and the gel relaxes to the COOH-induced final height $h(\infty)\approx h(0)\,.$}
\label{schematic}
\end{figure}

One chemically responsive system with a deformation rate faster than $\tau^{-1}$ through a tunable transient osmotic imbalance is the polyacrylic acid (PAA) hydrogel~\cite{Korevaar2020}. Under neutral or basic pH, the PAA gel can arrest divalent copper ions Cu$^{2+}$ (or calcium ions Ca$^{2+}$ as a biological signal mediator~\cite{Korevaar2020, VELAZQUEZ20111252, Mahadevan2000}) and contract with respect to its equilibrium height $H$ by forming COO$^--$Cu$^{2+}-$COO$^-$ chelates that remain kinetically stable without external Cu$^{2+}$  (Fig.~\ref{schematic}a)~\cite{Palleau2013}. When HCl is delivered as a second stimulus, the dissolved acid rapidly displaces Cu$^{2+},$ releasing it to the fluid phase of the gel (Fig.~\ref{schematic}b). Although the formation of the stable carboxyl (COOH) groups (this time in an acidic condition) favors gel contraction~\cite{Longo2014, Longo2016, Drozdov2016}, the gel temporarily swells by $\sim$10$\%$ of $H$ over the total copper decomplexation time $\tau_{total}$ if $\tau_{total}<\tau\equiv H^2/D$ ($D:$ poroelastic diffusion constant). When the Cu$^{2+}$ concentration equilibrates between the gel and the initially copper-free supernatant domain, the gel contracts to the height favored by the carboxyl groups (Fig.~\ref{schematic}c). As a control experiment, adding CuSO$_4$ into the HCl solution suppressed the swelling, implying a reduction of the gel hypotonicity due to the temporary free copper gradient in the first place~\cite{Korevaar2020} (Sec.~S1 in~\cite{si_dimensionless_eq}).

In this Letter, we theoretically address the following problem: Since osmosis is associated with interface selectivity to a solute, how does osmotic solvent influx into the gel emerge and then diminish across the gel-supernatant interface that is fully permeable to the ions and acid? We hypothesize that this influx is ``diffusio-osmotic" (Fig.~\ref{schematic}b): Steric interactions between the proton-doped polymer network and the free copper ions could lead to an interfacial tension $\gamma_{int}>0$ proportional to the free ion volume fraction $\phi^{(0)}$ along the polymer-fluid interface. As with Marangoni flows, the solvent must then flow towards higher $\gamma_{int}\,,$ defined as diffusio-osmosis~\cite{DERJAGUIN1993138, Marbach2019}. Momentum conservation demands that the diffusio-osmotic solvent velocity $\mathbf{v}_{DO}$ be counteracted by the polymer displacement with a velocity $\mathbf{v}_{DP}=-\mathbf{v}_{DO}$ (Fig.~\ref{schematic}b). In analogy with colloidal and polymer diffusio-phoresis~\cite{Marbach2019, Hinestrosa2020}, we define this reverse motion of the gel backbone as ``gel diffusio-phoresis," which transduces the diffusio-osmotic solvent flow into poroelastic deformations. To test this hypothesis, we develop a linear poroelastic theory for diffusio-phoretic gel swelling caused by the released copper gradient (Fig.~\ref{schematic}b). Previously, hydrogels were used as solute beacons to prolong the diffusio-phoretic migration times and distances of colloidal particles, albeit without coupling the gel mechanics and solute dynamics~\cite{Banerjee2016, Banerjee2020}. Furthermore, although polyacrylate and polyacrylamide hydrogels can transiently deform in response to osmolytes as heavy as 20--200~kDa, this was found to be due to the suppressed osmolyte diffusion in the gel~\cite{Sleeboom2017, Aangenendt2020, Punter2020, Tong1996}. Here we show how adding strong acid yields a diffusio-phoretic swelling burst with a rate $\tau_{total}^{-1}>\tau^{-1}\,,$ in quantitative agreement with experiments. For weak acid, swelling is suppressed  ($\tau_{total}^{-1}\lesssim\tau^{-1}$). Our theory confirms that molecules ($\lesssim 100$ Da) much smaller than typical osmolytes can induce diffusio-osmotic stress and deform the gel rapidly without impeded diffusion or interface selectivity. 

%Moreover, .

We formulate the gel mechanics via a minimal Biot consolidation model~\cite{Sachs1994, Biot1941}. Our model couples the gel mechanical response to the flow and concentration of the copper and acid as well as to the stresses they induce through association/dissociation with the gel backbone. In the gel domain, we use the incompressibility condition and the Darcy's law for porous flow to determine the fluid velocity relative to the solid matrix $\mathbf{v}$ and the solution pressure $p$  ($\mu_f:$ kinematic viscosity)~\cite{si_dimensionless_eq}, i.e.,
\begin{equation}
\nabla\cdot\left(\mathbf{v}+\frac{\partial \mathbf{u}}{\partial t}\right)= 0\,,\quad
\mathbf{v} = -\frac{k_f}{\mu_f}\nabla p\,. \label{eq:gelfluid}
\end{equation}
The matrix displacement vector $\mathbf{u}$ or equivalently the elastic strain tensor $\boldsymbol{\epsilon}\equiv (\nabla \mathbf{u} + (\nabla \mathbf{u})^T)/2$ is determined from the mechanical equilibrium condition for the gel stress tensor $\boldsymbol\sigma$
\begin{equation}
   \nabla\cdot\boldsymbol{\sigma}=0\,.
   \label{eq:solid1}
\end{equation}
In $\boldsymbol{\sigma}\,,$ the linear poroelastic terms comprise the elastic stresses ($\mu\,,\lambda\,:$ Lam\'{e} coefficients) and the solution pressure $p\,.$ We add to these two contractility terms associated with the bound ions ($\tilde{\gamma}\,,\tilde{\chi}:$ stress prefactors), the osmotic pressure induced by the polymer volume fraction $\phi_p\,,$ and the diffusio-osmotic term due to the interstitial free copper volume fraction $\phi^{(0)}$ ($\mathbf{I}:$ rank-two identity tensor, $k_B:$ Boltzmann's constant, $T\,:$ temperature, $v_c:$ molecular volume)~\cite{si_dimensionless_eq}:
\begin{equation}
\begin{split}
    \boldsymbol{\sigma} =2\mu \boldsymbol{\epsilon}+\mathbf{I}\Bigl[&\lambda \Tr\left(\boldsymbol{\epsilon}\right)-p \Bigr.+\tilde{\gamma} \phi^{(b)}+\tilde{\chi}\phi_+^{(b)}\Bigl.\\& -\frac{k_B T}{v_c}\biggl(\eta_{DP}\phi^{(0)}+\frac{\phi_p^2}{2}\biggr)\Bigr]\,.
\end{split}      
\label{eq:stress_tensor}
  \end{equation}
The dimensionless diffusio-phoretic coefficient satisfies $\eta_{DP}>0$ ($\eta_{DP}<0$) for repulsive (attractive) interactions between the polymer and the free copper. We consider steric repulsions with an exclusion radius $R_e$ that leads to $\eta_{DP}\equiv R_e^2/k_f>0\,,$ causing gel swelling when the free copper gradient is in the $-\mathbf{\hat{z}}$ direction (Fig.~\ref{schematic}b). From Eqs.~\ref{eq:gelfluid},~\ref{eq:stress_tensor}, the diffusio-phoretic velocity is defined as $\mathbf{v}_{DP}\equiv-D_{DP}\nabla\phi^{(0)}$ where $D_{DP}\equiv k_B T R_{e}^2/v_c \mu_f$ is the mobility~\cite{si_dimensionless_eq}. We ignore a similar effect stemming from the acid gradient since acid equilibrates across the two domains much faster than the timescales in this work. As with polymer solutions, the cross-linked polymers impose a permanent osmotic stress~\cite{DoiSoftMatter, hong2008}.  Because $D_{DP}\gg D$ holds, and the diffusio-osmotic stress linear in $\phi^{(0)}$ in general dominates the polymeric osmotic stress quadratic in $\phi_p\,,$ controlled ionic release from the PAA gel can enable very high strain and strain rates compared to mere osmotic solvent absorption.  

The internal gel stresses (Eq.~\ref{eq:stress_tensor}) are governed by the advection and diffusion of the free copper ($\phi^{(0)}$) and the free acid ($\phi^{(0)}_+$), as well as their conversion rates to/from the bound states $\phi^{(b)}\,,\phi^{(b)}_+$ on the gel backbone. They altogether satisfy $\phi_s+\phi_p+\phi^{(0)}+\phi_+^{(0)}+\phi^{(b)}+\phi^{(b)}_+=1$ ($\phi_s:$ solvent volume fraction). Our model captures the evolution of $\phi^{(0)}$ and $\phi^{(0)}_+$ through the reaction-transport equations ($D_x:$ diffusion constant of species $x\,,$ $\tilde{r}:$ rate constant, $\phi^\ast:$ COO$-$ volume fraction; Table~S1 in~\cite{si_dimensionless_eq})
\begin{equation}
\begin{split}
\frac{\partial \phi^{(0)}}{\partial t}+\nabla\cdot&\overbrace{\left[\phi^{(0)}\left(\mathbf{v}+\frac{\partial\mathbf{u}}{\partial t}\right) - D_{Cu} \nabla\phi^{(0)} \right]}^{\equiv \mathbf{Q}_{Cu}}\\ &=\underbrace{\tilde{r}\phi^{(0)}_+\phi^{(b)}-\tilde{r}\phi^{(0)}\left[\phi^\ast-2\phi^{(b)}-\phi^{(b)}_+\right]}_{\equiv R_{Cu}}\,,
\end{split}    
\label{eq:copper_RD}
\end{equation}
\begin{equation}
\begin{split}
\frac{\partial \phi^{(0)}_+}{\partial t}+\nabla\cdot &\overbrace{\left[\phi^{(0)}_+\left(\mathbf{v}+\frac{\partial\mathbf{u}}{\partial t}\right) - D_+ \nabla\phi^{(0)}_+ \right]}^{\equiv \mathbf{Q}_+}\\ &=-\underbrace{\tilde{r}\phi^{(0)}_+(\phi^\ast-\phi^{(b)}_+)}_{\equiv R_+}\,,
\end{split}\label{eq:acid_RD}
\end{equation}
where the flux terms $\mathbf{Q}_{Cu}\,, \mathbf{Q}_+$ involve particle advection with the flow and diffusion. The first term of the rate $R_{Cu}$ describes the acid-induced Cu$^{2+}$ release from the gel backbone, and the second term is the formation rate of a COO$^--$Cu$^{2+}-$COO$^-$ chelate. The source term $R_+$ is the COOH formation rate. Then, $\phi^{(b)}\,, \phi^{(b)}_+$ are determined by the rate equations
\begin{equation}
\frac{\partial \phi^{(b)}}{\partial t}= - R_{Cu}\,,\quad \frac{\partial \phi^{(b)}_+}{\partial t}= R_+\,.\label{eq:rate}
\end{equation}
For simplicity, we assume a single rate constant $\tilde{r}$ for all reactions in Eqs.~\ref{eq:copper_RD}--\ref{eq:rate} since they are found to be comparable in the experiments (Sec.~S2 and Fig.~S1 in~\cite{si_dimensionless_eq}). 

In the supernatant domain, denoting the fluid velocity by $\mathbf{V}$, the stress tensor by $\boldsymbol{\sigma}^{(a)}\,,$ and the pressure by $P\,,$ the incompressibility condition and the Stokes flow are 
\begin{equation}
\nabla\cdot\mathbf{V} = 0 \,,\quad \nabla\cdot\boldsymbol{\sigma}^{(a)} = 0\,; \quad \boldsymbol{\sigma}^{(a)} = \mu_f \nabla \mathbf{V}- \mathbf{I} P\,. \label{eq:supernatant_fluid}
\end{equation}
The copper ions with a volume fraction $\phi^{(a)}$ and acid with a volume fraction $\phi^{(a)}_+$ in the supernatant domain undergo only advection and diffusion with the fluxes $\mathbf{Q}_{Cu}^{(a)}\,, \mathbf{Q}^{(a)}_+$, governed by the mass conservation equations ($D^{(a)}_x:$ diffusion constant of species $x$ in the supernatant)~\cite{si_dimensionless_eq}
\begin{equation}
\frac{\partial \phi^{(a)}}{\partial t}+\nabla\cdot \overbrace{\left[\phi^{(a)}\mathbf{V} - D^{(a)}_{Cu} \nabla\phi^{(a)} \right]}^{\equiv \mathbf{Q}_{Cu}^{(a)}}=0\,,  
\label{eq:copper_supernatant}
\end{equation}
\begin{equation}
\frac{\partial \phi^{(a)}_+}{\partial t}+\nabla\cdot \overbrace{\left[\phi^{(a)}_+ \mathbf{V} - D^{(a)}_+ \nabla\phi^{(a)}_+ \right]}^{\equiv \mathbf{Q}^{(a)}_+}=0\,.
\label{eq:acid_supernatant}
\end{equation}

%Eqs.~\ref{eq:gelfluid}--\ref{eq:rate} constitute a set of nonlinear differential equations eighth order in space and fifth order in time. They are coupled to Eqs.~\ref{eq:supernatant_fluid}--\ref{eq:acid_supernatant}, which are seventh order in space and second order in time, through the continuity conditions between the two domains, given by ($\mathbf{\hat{n}}\,:$ surface normal vector)

Next, we determine the boundary conditions. The nonlinear differential equations~\ref{eq:gelfluid}--\ref{eq:rate} in the gel domain are coupled to Eqs.~\ref{eq:supernatant_fluid}--\ref{eq:acid_supernatant} in the supernatant domain through the following continuity conditions between the two domains ($\mathbf{\hat{n}}\,:$ surface normal)
\begin{equation}
\begin{split}
    z= H\,:&\quad \mathbf{V}=\mathbf{v}+\frac{\partial \mathbf{u}}{\partial t},\; P-p=\frac{k_B T}{v_c} \bigl(\eta_{DP}\phi^{(0)}+\frac{\phi_p^2}{2}\bigr)\,,\\& \mathbf{\hat{n}}\cdot\boldsymbol{\sigma}=\mathbf{\hat{n}}\cdot\boldsymbol{\sigma}^{(a)}\,, \quad \mathbf{\hat{n}}\cdot\mathbf{Q}_{Cu}=\mathbf{\hat{n}}\cdot\mathbf{Q}_{Cu}^{(a)}\,,\\& \mathbf{\hat{n}}\cdot\mathbf{Q}_+=\mathbf{\hat{n}}\cdot\mathbf{Q}_+^{(a)}\,,\quad \phi^{(0)}=\phi^{(a)}\,,\quad \phi^{(0)}_+=\phi^{(a)}_+\,.
\end{split}
    \label{eq:BC1}
\end{equation}
The interfacial pressure jump $P-p$ is set by the polymer-induced osmotic stress that relaxes the gel to its equilibrium state (without copper and acid), which we take as the reference state with zero strain. Adding a diffusio-osmotic agent (copper) will alter the gel solution pressure in the gel, which must equilibrate instantaneously across the interface~\cite{hong2008, hong2009}, leading to the second condition in Eq.~\ref{eq:BC1}. The boundary conditions for the gel attached to a rigid, impermeable substrate at $z=0$ and for the impermeable supernatant domain boundary at $z=H+H^{(a)}$ are given by ($H^{(a)}:$ supernatant domain height)~\cite{si_dimensionless_eq}
\begin{equation}
\begin{split}
    z=0\,:&\quad \mathbf{\hat{n}}\cdot\mathbf{v}=0\,,\quad \mathbf{u}=0\,,\\& \mathbf{\hat{n}}\cdot\mathbf{Q}_{Cu}=0\,,\quad \mathbf{\hat{n}}\cdot\mathbf{Q}_+=0\,,
    \end{split}
    \label{eq:BC2}
\end{equation}
\begin{equation}
\begin{split}
    z=H+H^{(a)}\,:&\quad \mathbf{V}=0\,,\quad P=0\,,\\&\mathbf{\hat{n}}\cdot\mathbf{Q}_{Cu}^{(a)}=0\,,\quad \mathbf{\hat{n}}\cdot\mathbf{Q}_+^{(a)}=0\,.
\end{split}
    \label{eq:BC3}
\end{equation}

To explain the vertical deformation dynamics in Ref.~\cite{Korevaar2020}, we consider first, 1D uniaxial deformations in response to a uniform acid front advancing in the $-\mathbf{\hat{z}}$ direction to the copper-laden gel and, second, 2D deformations due to an acid front with a Gaussian weak perturbation to investigate the effect of the potential nonuniformities during acid delivery in the experiments. In the linear elastic limit, we take the polymer volume fraction $\phi_p$ and the COOH volume fraction $\phi^\ast$ constant by ignoring the effect of small deformations on the concentrations~\cite{si_dimensionless_eq}. In 1D (in the $\pm\mathbf{\hat{z}}$ direction), we denote the magnitudes of all the vectors by $u_z (z, t)\equiv |\mathbf{u}|\,, v(z,t)\equiv|\mathbf{v}|\,, V(z,t)\equiv|\mathbf{V}|\,.$ All non-vanishing tensors reduce to a scalar, i.e., $\sigma_{zz}\equiv\mathbf{\hat{z}}\cdot\boldsymbol{\sigma}\cdot\mathbf{\hat{z}}\,,$ $\epsilon_{zz}\equiv\mathbf{\hat{z}}\cdot\boldsymbol{\epsilon}\cdot\mathbf{\hat{z}}=\partial u_z/\partial z\,,$ $\mathbf{\hat{z}}\cdot\mathbf{I}\cdot\mathbf{\hat{z}}=1\,.$ Eqs.~\ref{eq:gelfluid}--\ref{eq:BC3} determine the uniaxial deformations as follows: Per Eqs.~\ref{eq:BC1}--\ref{eq:BC3}, the incompressibility conditions in Eqs.~\ref{eq:gelfluid},~\ref{eq:supernatant_fluid} reduce to $v+\partial u_z/\partial t=0$ and $V=0\,,$ i.e., local gel deformations do not impose any net flow in the lab frame. This also leads to a diffusive stimulus dynamics in Eqs.~\ref{eq:copper_RD},~\ref{eq:acid_RD},~\ref{eq:copper_supernatant}, and~\ref{eq:acid_supernatant}. Then, using the unitless variables $u_z'\equiv u_z/H\,, z'\equiv z/H$ (gel), $z'\equiv z/H^{(a)}$ (supernatant), $ t'\equiv t/\tau$ where $\tau\equiv\mu_f H^2/k_f \bar{p}\,,$ $\bar{p}\equiv (2\mu+\lambda)\,,$ and dropping their primes, Eqs.~\ref{eq:gelfluid}--\ref{eq:stress_tensor} yield a dimensionless evolution equation for the gel displacement as ($\gamma\equiv\tilde{\gamma}/\bar{p}\,, \chi\equiv\tilde{\chi}/\bar{p}\,, \nu_{DP}\equiv k_B T \eta_{DP}/v_c\bar{p}$)~\cite{si_dimensionless_eq}
\begin{equation}
        \frac{\partial u_z}{\partial t}=\frac{\partial^2 u_z}{\partial z^2} + \gamma \frac{\partial \phi^{(b)}}{\partial z} + \chi \frac{\partial \phi^{(b)}_+}{\partial z}- \nu_{DP}\frac{\partial\phi^{(0)}}{\partial z}
    \label{eq:1D_model}
\end{equation}
with the boundary conditions from Eqs.~\ref{eq:BC1},~\ref{eq:BC2}  ($\mathbf{\hat{n}}=\mathbf{\hat{z}}$)
\begin{equation}
u_z\big|_{z=0}=0\,,\quad\frac{\partial u_z}{\partial z}\bigg|_{z=1}=- \gamma\phi^{(b)} - \chi\phi^{(b)}_+\,. 
\label{eq:1D_BC}    
\end{equation}
Eqs.~\ref{eq:1D_model},~\ref{eq:1D_BC} are closed by the unitless forms of Eqs.~\ref{eq:copper_RD}--\ref{eq:rate},~\ref{eq:copper_supernatant},~\ref{eq:acid_supernatant} (i.e., Eqs.~S1-S5; Sec.~S2) and the corresponding boundary conditions in Eqs.~\ref{eq:BC1}--\ref{eq:BC3} with the unitless parameters in Table~S1~\cite{si_dimensionless_eq}. The seven initial conditions for the contracted gel with stored Cu$^{2+}$ are 
\begin{subequations}
\begin{equation}
u_z=-\gamma\phi^{(b)}z\,, \phi^{(b)}=\frac{\phi^\ast}{2}\,, \phi^{(b)}_+=\phi^{(0)}=\phi^{(0)}_+=0\,,
    \label{eq:IC1}
\end{equation}
and in the supernatant solution
\begin{equation}
\phi^{(a)}=0\,, \phi^{(a)}_+(z)= \frac{\phi^{(a)}_{+,i}}{2}\left[1+\tanh\left(\Gamma (z-z_0)\right)\right]\,,
    \label{eq:IC2}
\end{equation}
\end{subequations}
where $\Gamma = \Gamma^{(1\text{D})}$ and $z_0=z_0^{(1\text{D})}$~\cite{si_dimensionless_eq}.

\begin{figure}[ht!]
\includegraphics[width=1\columnwidth]{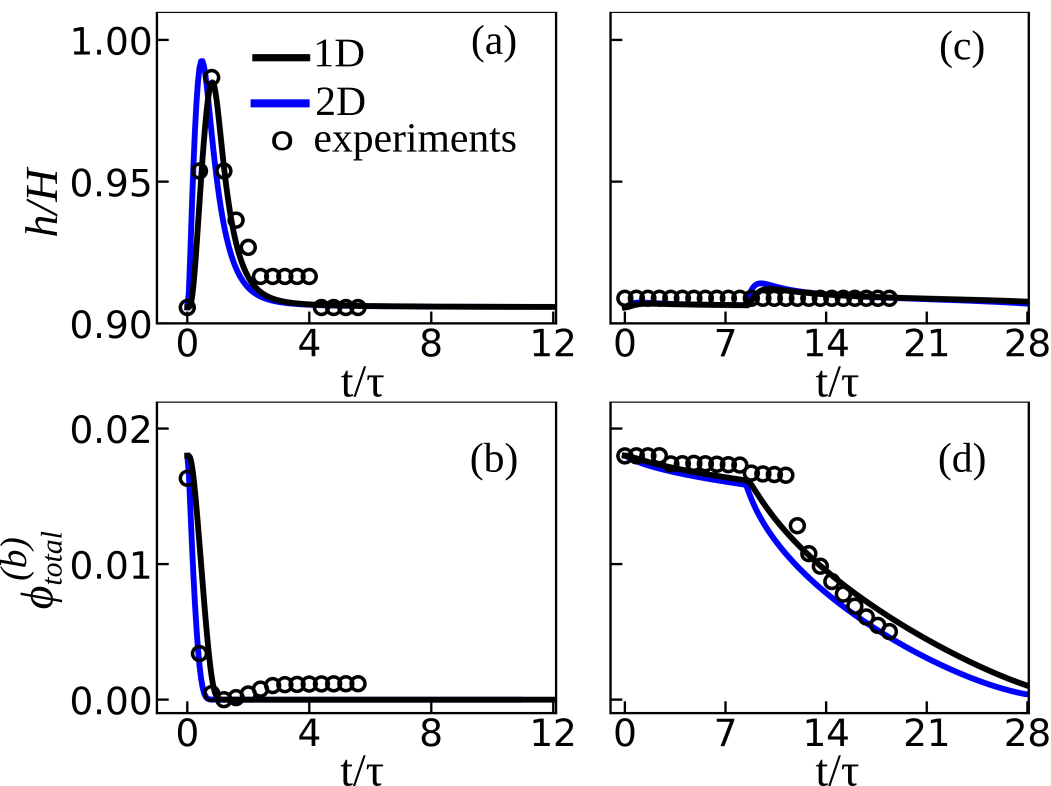}
\caption{\textbf{Gel deformations upon acid addition.} The gel height $h/H\equiv 1+u_z$ versus time for \textbf{(a)} $\phi_{+,i}^{(a)}=0.006$ (equivalent to $\sim$1M acid) and \textbf{(c)} the step-wise addition of acid with $\phi_{+,i}^{(a)}=6\times 10^{-5}$ ($\sim$0.01M) initially and $\phi_{+,i}^{(a)}=3\times 10^{-4}$ ($\sim$0.05M) at $t/\tau=8.4$, respectively (see Eq.~\ref{eq:IC2}). For the 2D simulations, the hydrogel height is calculated by averaging over the experimental distance between two adjacent microplates ($=5$ $\mu m$) about the horizontal center of the gel film ($x/L=0.5\pm 1\times10^{-4}$). \textbf{(b)}, \textbf{(d)} Time dependence of the total bound copper $\phi^{(b)}_{total}\equiv\int^1_0 \phi^{(b)} dz$ corresponding, respectively, to (a) and (c). In 2D, $\phi^{(b)}_{total}$ is averaged over $x/L=0.5\pm 1\times10^{-4}$ as with the gel height.}
\label{fig:results1}
\end{figure}

\begin{figure}
\includegraphics[width=1\columnwidth]{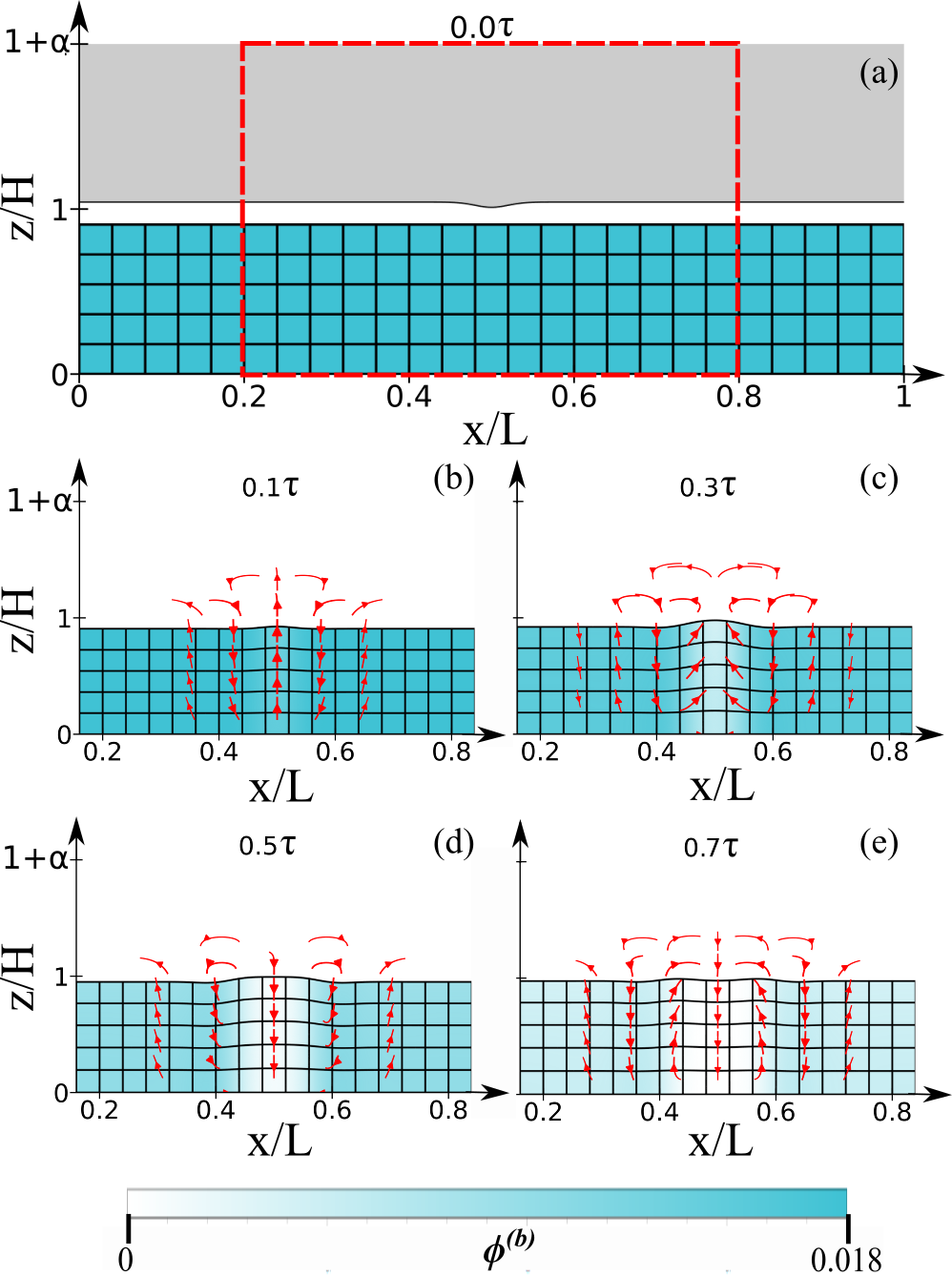}
\caption{\textbf{2D gel response to a 1M acid stimulus.} \textbf{(a)} For 1M acid with a Gaussian perturbation (Eqs.~\ref{eq:IC2},~S6), the gel dynamics at \textbf{(b)} $t=0.1\tau$, \textbf{(c)} $t=0.3\tau$, \textbf{(d)} $t=0.5\tau$ and \textbf{(e)} $t=0.7\tau$ within the boxed region shown in (a). The red streamlines indicate the computed fluid flow in the lab frame (line width: logarithm of the flow speed, arrows: flow direction). The blue color scale indicates the bound copper volume fraction ($\alpha=770$)~\cite{si_dimensionless_eq}. } \label{fig:results2_strong}
\end{figure}

Our main results are demonstrated in Fig.~\ref{fig:results1}. The simulation procedure and post-processing of the experimental data are detailed in Sec.~S3, and the validity of the linear poroelastic model is discussed in Sec.~S4~\cite{si_dimensionless_eq}. Upon the diffusion of 1M acid ($\phi^{(a)}_{+, i}=0.006$) into the copper-laden gel from the supernatant domain, the gel height exhibits a temporal spike with a magnitude $\lesssim0.1 H$ in quantitative agreement with experiments (Fig.~\ref{fig:results1}a). This rapid swelling followed by the slower contraction can be understood by considering the interplay among the acid flux and its complexation, the subsequent release of bound copper and the diffusio-osmotic solvent inflow induced by it, and the gel relaxation at longer times (Fig.~S2-S4)~\cite{si_dimensionless_eq}: Although our theory suggests that diffusio-phoresis can induce rapid gel deformations at a timescale $\tau/\nu_{DP}\ll \tau\,,$ the swelling rate is limited by the overall release time $\tau_{total}\approx 0.82\tau$ of the height-averaged bound copper $\phi^{(b)}_{total}$ (Fig.~\ref{fig:results1}b). As a result, the gel undergoes continual diffusio-phoretic swelling until $t\approx \tau_{total}$ when the height reaches maximum (Fig.~\ref{fig:results1}a). This swelling time is still shorter than $\tau$ and can be improved by considering nonlinear deformations driven by higher acid concentrations. The decay from the maximum height is governed by the competition between the diffusive relaxation of the deformations and the residual diffusio-phoretic swelling. This leads to a subdiffusive relaxation with a timescale $\tau_{r, 1}\approx~0.54\tau\,,$ which is higher than the timescale $\tau_D\equiv 4 \tau/\pi^2 \approx 0.4 \tau$ of the purely diffusive relaxation dynamics (Sec.~S5, Fig.~S5,~S6)~\cite{si_dimensionless_eq}.  The diffusive contribution ceases at $t\gtrsim 4\tau\,,$ and the longtime slow relaxation is governed by the ever damping diffusio-phoretic term with a timescale $\tau_{r, 2}\approx 5.2\tau$~\cite{si_dimensionless_eq}.

To validate that the swelling in Fig.~\ref{fig:results1}a is driven by the rapid release of the Cu$^{2+}$ ions, the same acid amount was slowly added over successive steps with concentrations ranging from 0.01~M to 1~M, which lead to no deformation (Fig.~\ref{fig:results1}c, circles)~\cite{Korevaar2020}. Here we simulate only the first two steps with $\phi_{+,i}^{(a)}=6\times 10^{-5}$ ($\sim$0.01M) at $t=0$ and $\phi_{+,i}^{(a)}=3\times 10^{-4}$ ($\sim$0.05M) at $t=8.4\tau\,.$ Our numerical results yield marginal deformations about 0.1$\%$ and 0.5$\%$ of the equilibrium height $H$ that fall within the experimental error of $\pm 1\%~H$ (Fig.~\ref{fig:results1}c). Swelling is suppressed for the low acid concentrations since the bound Cu$^{2+}$ release rate is drastically reduced (Fig.~\ref{fig:results1}d). In this limit, the gel poroelastic relaxation balances the diffusio-phoretic swelling that is slowed down by the low bound Cu$^{2+}$ release rate. Consequently, the relaxation of the minute deformations is subdiffusive (Fig.~S6)~\cite{si_dimensionless_eq}. 

Although the 2D weakly perturbed gel swelling and bound copper release profiles deviate only slightly from the 1D uniaxial deformation results (Fig.~\ref{fig:results1}a--d, blue curves), the 2D simulations reveal the deformation dynamics during swelling and relaxation starting from the initial conditions given in Sec.~S3~\cite{si_dimensionless_eq}. When adding 1M acid initially (Fig.~\ref{fig:results2_strong}a), the penetration of the Gaussian stimulus front (with a standard deviation in the order of the water capillary length) into the gel triggers a local swelling bump associated with a convective flow (Fig.~\ref{fig:results2_strong}b--c, Movie~S1). The flow reverses direction at the center ($x=L/2$) at the onset of break-up of the single bump into two swelling fronts ($t=0.5\tau\,,$ Fig.~\ref{fig:results2_strong}d), which travel in opposite directions at the gel surface in phase with the copper decomplexation front ($t=0.7\tau\,,$ Fig.~\ref{fig:results2_strong}e) and decay at longer times along with the diminishing flow streamlines. Similar traveling deformation fronts sensitive to the acid progression rate and direction were reported in Ref.~\cite{Korevaar2020}. For 0.05 M acid delivery at $t=8.4\tau$ after the initial 0.01M acid addition step, because the deformation and flow are negligible, a traveling front at the gel surface does not form (Fig.~S7, Movie~S2)~\cite{si_dimensionless_eq}.

Our theory explains the fast swelling dynamics of the PAA gel by introducing a gel diffusio-phoresis mechanism, which nevertheless needs to be validated by microscopic approaches such as molecular dynamics simulations. Furthermore, the linear poroelastic swelling in Fig.~\ref{fig:results1}a only produces a strain rate of $\sim 0.04~$s$^{-1}$ and a power density of $\sim 11.1$~mW/kg (Sec.~S6, Fig.~S8), which are surpassed by certain gels that generate 0.2 s$^{-1}$ and $260$ mW/kg and PAA microgel suspensions that achieve $230$ mW/kg through the osmotic swelling of dry polymers exposed to a solvent~\cite{si_dimensionless_eq, Arens2017, Choudhary2022}. Therefore, our analysis must be extended to nonlinear large deformations to test high strain rates and power densities based on $D_{DP}\gg D$ allowed by the gel diffusio-phoresis~\cite{hong2008, MacMinn2016}. This inequality can also be engineered in other gels to scale up chemically responsive shape-shifting hydrogel actuators. Also, hydrogels that combine diffusio-phoresis with periodic actuations via cyclic chemical feedback (as those in Belousov–Zhabotinsky gels~\cite{Balazs2013, epstein2014}) must be designed for engineering applications. These steps will pave the way for internally powered gel-based proof-of-concept soft robots with enhanced precision, versatility, and dexterity. 
\\

%TC:ignore

\begin{acknowledgments}
We thank J. Aizenberg, J. Barone, S. Cheng, J. Gray, A. Grinthal, M. Pleimling, W. Shu for fruitful discussions and the Virginia Tech College of Science for financial support. We acknowledge the Virginia Tech Advanced Research Computing Center for computing resources.
\end{acknowledgments}

%TC:endignore

%\bibliography{apssamp} 

\providecommand{\noopsort}[1]{}\providecommand{\singleletter}[1]{#1}%
%

%%%%%%%%% Supplementary information

\onecolumngrid
\newpage

\begin{center}
  \textbf{\large Supplementary information for:\\ Diffusio-phoretic fast swelling of chemically responsive hydrogels}\\[.2cm]
  Chinmay Katke$^{1, 2\,}$, Peter A. Korevaar$^{3\,}$, and C. Nadir Kaplan$^{1, 2*}$\\[.1cm]
  {\itshape ${}^1$Department of Physics,
 Virginia Polytechnic Institute and State University,
 Blacksburg, VA 24061, USA,\\
  ${}^2$Center for Soft Matter and Biological Physics, Virginia Polytechnic\\ 
  Institute and State University,
Blacksburg, VA 24061, USA,\\
  ${}^3$Institute for Molecules and Materials,
Radboud University, 6525 AJ Nijmegen, The Netherlands.\\}
  ${}^*$\url{nadirkaplan@vt.edu}\\
\end{center}

\setcounter{equation}{0}
\setcounter{figure}{0}
\setcounter{table}{0}
\setcounter{page}{1}
\setcounter{section}{0}
\renewcommand{\theequation}{S\arabic{equation}}
\renewcommand{\thefigure}{S\arabic{figure}}
\renewcommand{\thetable}{S\arabic{table}}

\renewcommand{\thesection}{S\arabic{section}}

All source codes pertaining to the 1D and 2D simulations reported in this paper and Movies S1, S2 can be downloaded at
\begin{center}
  \url{https://github.com/nadirkaplan/hydrogel_diffusiophoresis}.\\   
\end{center}

\vskip 0.5in

\noindent
{\bf Movie S1.} The hydrogel matrix displacement $\mathbf{u}$ (mesh grid), the unitless fluid flow speed $U/U_0$ (color scale; velocity scale $U^{(0)}\equiv H/\tau=4\times 10^{-6} m/s$) and flow direction (red arrows) are depicted for the addition of 1M acid. The line width of the red arrows is proportional to the natural logarithm of the unitless flow speed. 
\\

\noindent
{\bf Movie S2.} The hydrogel matrix displacement $\mathbf{u}$ (mesh grid), the unitless fluid flow speed $U/U_0$ (color scale; velocity scale $U^{(0)}\equiv H/\tau=4\times 10^{-6} m/s$) and flow direction (red arrows) are depicted for the addition of 0.01M acid at $t=0$ and the subsequent addition of 0.05M acid at $t=8.4\tau\,.$ The line width of the red arrows is proportional to the natural logarithm of the unitless flow speed.

\section{Origins of osmosis-induced gel swelling upon copper release}

When Cu$^{2+}$ is released from the PAA gel backbone upon acid addition, the gel crosslinking density also changes since the COO$^--$Cu$^{2+}-$COO$^-$ chelates are broken. This potentially enables gel expansion (until pH-reduction induced contraction takes place) according to the Flory-Rehner theory that links swelling to the crosslinking density~\cite{FloryRehner}. In this section we explain why this alternate mechanism is not applicable since it suggests a much slower swelling dynamics than that observed in our experiments, and how our control experiments in Ref.~\cite{Korevaar2020_2} validate that the gel swelling is dominated by the osmotic imbalance generated by the released copper.

\begin{enumerate}
    \item \underline{Control experiments validate osmosis-induced swelling due to released copper ions.} Experiments reported in the Supplementary Fig. 3 of Ref.~\cite{Korevaar2020_2} demonstrated that the transient swelling was suppressed when Cu$^{2+}$ was also included in the HCl solution by adding CuSO$_4$. Using calcium (Ca$^{2+}$) instead of Cu$^{2+}$ led to similar results (Supplementary Fig. 4, Ref.~\cite{Korevaar2020_2}). Based on these observations, we had concluded in that the transient swelling responses are induced by the osmotic pressure change due to free Cu$^{2+}$ ions in the hydrogel, and this effect is diminished when Cu$^{2+}$ ions added to the HCl solution reduce the hypotonic character of the gel.

    \item \underline{Potential effect of chelation suggests slow swelling and fast de-swelling in contradiction with experiments.} The poroelastic timescale is given by $\tau\equiv \mu_f H^2/k_f \bar{p}$ (Table~\ref{table:simulation_parameters}). If the modulation of the crosslinking density controls the transient gel swelling and de-swelling, then the product of the hydraulic permeability and the elastic modulus $k_f \bar{p}$ must change upon chemical complexation/decomplexation. We will denote this quantity for the copper-laden gel as $\beta_0\equiv k_{f0} \bar{p}_0\,,$ for the copper-decomplexed gel as $\beta_1\equiv k_{f1} \bar{p}_1\,,$ and for the acid-complexed gel as $\beta_2\equiv k_{f2} \bar{p}_2\,.$ Then, the change in crosslinking due to the released Cu$^{2+}$ implies $\beta_1<\beta_0$ during swelling, and the subsequent change in hydrophibicity at reduced pH upon acid delivery implies $\beta_1<\beta_2$ during de-swelling. Note that a comparison between $\beta_0$ and $\beta_2$ cannot be devised without a microscopic theory. The inequality $\beta_1<\beta_2$ leads to $\tau_1>\tau_2\,,$ which means that swelling rate must be lower than the de-swelling rate. However, in the experiments swelling is faster than de-swelling (Fig.~\ref{timescales}c), meaning that the de-swelling rate is not governed by the formation rate of the COOH groups. For these reasons, the role of the structural changes in the gel backbone should not be relevant for the transient swelling/de-swelling observed in our experiments. We also note that the longtime slow relaxation beyond $t>=4\tau$ with a timescale $\tau_{r, 2}\approx 5.2 \tau$ governed by the ever damping diffusio-phoretic term (Fig.~\ref{timescales}a) is not captured in the experiments most likely because the variation in the height profiles at $t>=4\tau$ remains within the experimental error of $±1\%~H$.
    \item \underline{Lower elastic moduli at reduced crosslinking lead to much slower dynamics than in experiments.} For the pressure scale $\bar{p}=10^5~$Pa, we estimated the poroelastic deformation timescale as $\tau=2.5 s$ and found the swelling time to be $\tau_{total}=0.82 \tau \approx 2s$ in Fig. 2(a) (Table~\ref{table:simulation_parameters}). If the swelling were due to the reduction of the crosslinking density, the timescale would increase with decreasing modulus: In Ref.~\cite{Palleau2013_2} the PAA hydrogel fully complexed with copper via electrochemical delivery is measured to have an elastic modulus of 300 kPa as opposed to a gel without copper that has an elastic modulus of 21 kPa. This corresponds to a 15-fold decrease in the elastic modulus upon copper decomplexation, which would imply a much slower swelling process with $\tau_{total}~=30s$. Therefore, the fast gel swelling as a central result of this paper would not be observed in the experiments. Note that this analysis does not take into account the subsequent rapid acid complexation (at much higher rates than the mechanical deformation rates considered here) upon copper release. If the acid complexation is also considered, it is plausible to assume a constant elastic modulus. To this end, we took constant Lam\'{e} coefficients to keep the theory sufficiently simple while still accurately capturing the experimental measurements.
\end{enumerate}

%\subsection{}

\section{Unitless 1D reaction-transport equations of copper and acid}

Below, the reaction-diffusion dynamics of the free copper and acid in the gel is modeled by Eqs.~\ref{eq:copper_RD} and~\ref{eq:acid_RD}, respectively. The time-dependent volume fractions of the bound copper and acid are determined by Eqs.~\ref{eq:rate}. Within the supernatant solution above the gel, the dynamics of the two chemical stimuli are set by Eqs.~\ref{eq:copper_supernatant} and~\ref{eq:acid_supernatant}. The dimensionless parameters $\xi_{Cu}\,,~\xi_{+},~\xi^{(a)}_{Cu}\,,~\xi^{(a)}_+,~r$ are defined in Table~\ref{table:simulation_parameters}. The boundary conditions are given in Eqs.~10--12.

\begin{equation}
\frac{\partial \phi^{(0)}}{\partial t}= \xi_{Cu}\nabla^2\phi^{(0)}+r\phi^{(0)}_+\phi^{(b)}-r\phi^{(0)}\left[\phi^\ast-2\phi^{(b)}-\phi^{(b)}_+\right]\,,
\label{eq:copper_RD}
\end{equation}

\begin{equation}
\frac{\partial \phi^{(0)}_+}{\partial t}=\xi_{+}\nabla^2\phi^{(0)}_+ -r\phi^{(0)}_+(\phi^\ast-\phi^{(b)}_+)\,,
\label{eq:acid_RD}
\end{equation}

\begin{equation}
\frac{\partial \phi^{(b)}}{\partial t}= -r\phi^{(0)}_+\phi^{(b)}+r\phi^{(0)}\left[\phi^\ast-2\phi^{(b)}-\phi^{(b)}_+\right]\,,\quad \frac{\partial \phi^{(b)}_+}{\partial t}= r\phi^{(0)}_+(\phi^\ast-\phi^{(b)}_+)\,.\label{eq:rate}
\end{equation}

\begin{equation}
\frac{\partial \phi^{(a)}}{\partial t}=\xi^{(a)}_{Cu}\nabla^2 \phi^{(a)}\,,  
\label{eq:copper_supernatant}
\end{equation}
\begin{equation}
\frac{\partial \phi^{(a)}_+}{\partial t}=\xi^{(a)}_+\nabla^2\phi^{(a)}_+\,.
\label{eq:acid_supernatant}
\end{equation}

To explain the choice of a single rate constant $r$ in Eqs.~\ref{eq:copper_RD}--\ref{eq:rate}, we consider two outputs from the experiments~\cite{Korevaar2020_2}:
\begin{enumerate}
    \item {\it Deformation read-out:} The PAA hydrogel thin film contains an array of surface-attached, slightly pre-tilted epoxy microplates on an epoxy substrate. The tilting of plates enables real-time visualization of the micron-scale gel deformations.
    \item {\it Color read-out:} The complexation of the copper ions with the COO- groups turns the gel blue. The decomplexation of copper makes the gel colorless. 
\end{enumerate}

\begin{figure}[ht!]
\includegraphics[width=1\textwidth]{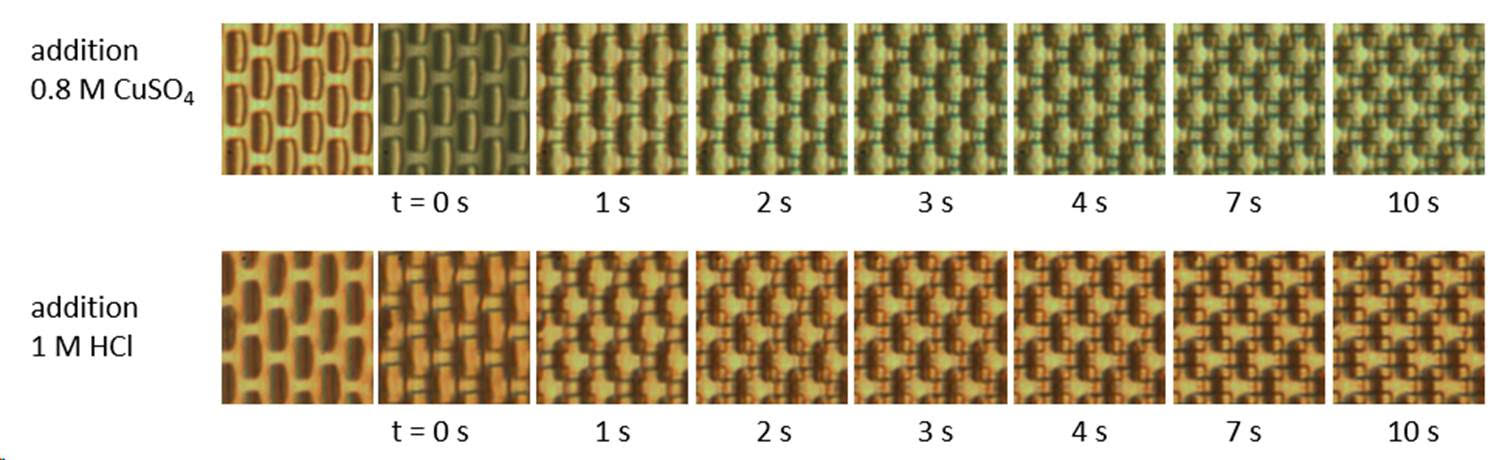}
\caption{{\bf Stimulus interactions with the PAA gel backbone.} {\bf Top row.} The association dynamics of Cu$^{2+}$ to the PAA hydrogel backbone when 0.8 M CuSO$_4$ is added to the PAA gel. {\bf Bottom row.} The COOH formation dynamics when 1 M HCl is added. In both cases, the starting condition is a PAA gel that is deprotonated and free of copper complex, and hence the microplates are in the upright orientation. The complexation dynamics of copper is extracted from the color change and gel deformations , i.e., the tilting of the plates. The COOH formation dynamics is extracted solely from the plate tilting patterns.}
\label{fig:rates}
\end{figure}

In Fig.~\ref{fig:rates} top row, the association of Cu$^{2+}$ to the gel occurs within the first 2 seconds ($t=0.8\tau\,;$ after that point no significant color change is observed), which informs the rate constant of the copper complexation (the rate constant of the last terms of Eqs.~\ref{eq:copper_RD} and of the first equation in Eq.~\ref{eq:rate}). After that, the gel contraction is finalized within 10 seconds ($=4\tau$). For the addition of HCl (Fig.~\ref{fig:rates}, bottom row), microplate tilting within the first 1-2 seconds implies that H$^{+}$ has associated to the gel within this time frame (this process does not change the gel color), which puts an upper bound to the rate constant of acid association (last terms of Eq.~\ref{eq:acid_RD} and of the second equation in Eq.~\ref{eq:rate}). The gel is fully contracted again within approximately 10 seconds. 
As for the copper decomplexation rate, the experimental bound copper profile in Fig. 2b indicates that copper is replaced by H$^+$ over the course of approx. 2 seconds. Subsequently, the gel contracts again (after the swelling pulse) within approx. 10 seconds. 

Because these three experimental datasets exhibit that the complexation-decomplexation processes occur within 1-2 seconds, it is plausible to assume that their rate constants are at the same order of magnitude. The rate constant $r$ is a fitting parameter, and we have chosen $r=1.25\times 10^4$ (or $\tilde{r}=5\times10^3 s^{-1}$ in real units). Since the reaction rates are much bigger than the diffusion and deformation rates at least by an order of magnitude, we have made use of this timescale separation to take the rate con-stants appearing Eqs.~\ref{eq:copper_RD}--\ref{eq:rate} equal to each other for simplicity.

Unless the reaction rates are much lower than the poroelastic relaxation rate, the diffusio-phoretic swelling effect will remain. In fact, this effect is suppressed upon stepwise addition of 0.01M and 0.05M acid in Fig. 2c, d exactly for this reason: The acid-dependent rate terms in Eqs.~\ref{eq:copper_RD}--\ref{eq:rate} are reduced, which brings the timescale of the overall complexation-decomplexation mechanism to be comparable to that of relaxation (any potential swelling is preempted by sufficiently rapid relaxation in this limit). This acid-induced control over the reaction rates enables the system to switch between rapid swelling (Fig. 2a) and negligible swelling (Fig. 2c).

\section{Simulation procedure and post-processing of experimental data \label{post_processing}}

 For uniaxial deformations, we solved Eqs.~13-15b and Eqs.~\ref{eq:copper_RD}--\ref{eq:acid_supernatant} by using the FEniCS finite element analysis (FEA) library on Python 3.9~\cite{AlnaesEtal2015}. 
 
 For weakly perturbed 2D dynamics, by making the horizontal coordinate $x$ unitless with the domain length $L\,,$ we complement the unitless forms of Eqs.~1--12 with periodic boundary conditions at $x=0$ and $x=1$ for the gel variables $\boldsymbol{f}\equiv\{p, \mathbf{v}, \mathbf{u}, \phi^{(0)}, \phi^{(0)}_+, \phi^{(b)}, \phi^{(b)}_+\}$ and $\boldsymbol{\sigma}\,,$ the supernatant domain variables $\boldsymbol{f}^{(a)}\equiv\{P, \mathbf{V}, \phi^{(a)}, \phi^{(a)}_+\}$ and $\boldsymbol{\sigma}^{(a)}$, as well as all flux terms. Eqs.~15a,~15b again hold with $\Gamma=\Gamma^{(2\text{D})}$ and $z_0=z_0^{(2\text{D})}$, which is given by a Gaussian profile, as well as with an additional initial condition for horizontal deformations
\begin{equation}
u_x\equiv \mathbf{u}\cdot\mathbf{\hat{x}}=0\,,\quad z_0^{(2\text{D})} =h_1 - h_2 e^{-\frac{(x-1/2)^2}{2\lambda^2}}\,.
\label{eq:IC3}
\end{equation}
Here $h_2$ is obtained from the initial supernatant acid amount constraint $\phi_{+, i}^{(a)}(2-z_0^{(1\text{D})})$ when $h_1$ is fixed, and $\lambda$ sets a perturbation width about the capillary length of water $\ell_c\sim1~$mm (see Fig.~3a for the initial acid profile). Table~\ref{table:simulation_parameters} lists the values of $h_1\,, h_2\,,\lambda\,,$ and $\Gamma^{(2\text{D})}\,.$
 
To obtain the 2D results, we solved Eqs.~1--12,~15a,~15b, and~\ref{eq:IC3} with periodic boundary conditions along the $x-$axis via the COMSOL Multiphysics 5.4 FEA package~\cite{Comsol}. The two FEA libraries yield the same result for uniform deformations (Fig.~\ref{fig:comparison}). All characteristic physical scales of the system and the unitless simulation parameters are listed in Table~\ref{table:simulation_parameters}. 

We convert the molar concentrations used in the experiments to the volume fractions used in our theoretical formulation as follows: Let $c$ be the molar concentration of the chemical species with the unit moles/liter. With molecular volume $v_c \approx 10^{-29}$m$^3$ and Avogadro's number $N_m \approx 6 \times 10^{23}\,,$ the volume of the solute with the molar concentration $c$ will be equal to $cv_cN_m = c \times 6 \times 10^{-6} $m$^3\,.$ Therefore, the volume fraction corresponding to $c$ molar solution is $cv_cN_m/1$L $= c \times 6 \times 10^{-3}$ ($1$L $= 10^{-3}$m$^3$). Then, $c^{(a)}_{+,i}=0.01$M, $0.05$M and $1$M of acid in the supernatant solution correspond to volume fractions of $\phi^{(a)}_{+,i}=6\times10^{-5}$, $3\times10^{-4}$ and $6\times10^{-3}\,,$ respectively~\cite{Korevaar2020_2}. The maximum concentration of the bound copper was estimated to be $c^{(b)}=2.9$M based on the absorption spectra of the ethylenediethylaminetetraacetate (EDTA) solution used to extract the bound copper from the hydrogel~\cite{Korevaar2020_2}. Thus, the volume fraction corresponding to the initial bound copper concentration is $\phi^{(b)}(z,t=0) \approx 0.018\,.$ Furthermore, assuming that all the carboxylate groups are complexed at $t=0$, the volume fraction of the COO$-$($\phi^*$) would be twice that of $\phi^{b}(z,t=0)$ giving $\phi^{\ast} = 0.036$ since each chelate is composed of two COO$-$ groups and one Cu$^{2+}$ ion. The acrylic acid concentration for the hydrogel is estimated to be $c_p = 6.7$M using the volume and concentration of the each component in the precursor solution to the hydrogel\cite{Korevaar2020_2}, which corresponds to the polymer volume fraction $\phi_p\approx0.04$.

To compare our simulations for a 1D uniform acid front and weakly perturbed 2D acid front with experiments, we used the uniform acid delivery and uniaxial deformation data illustrated in Fig.~3 of Ref.~\cite{Korevaar2020_2}. For uniaxial deformations (i.e., no $x-$dependence), we extracted the experimental time-dependent gel height $h(t)$ from the tilt angle data of the microplate array patterned into the gel film for real-time deformation readout~\cite{Korevaar2020_2}.  Defining the time-dependent microplate tilt angle as $\theta(t)\,,$ the unitless hydrogel height $h(t)$ is given by
\begin{equation}
    h(t) = \frac{\cos\theta(t)}{\cos\theta_0}\,, \label{eq:height}
\end{equation}
where $\theta_0=9^o$ in the native state of the gel without any complexed ions. At $t=0\,,$ the microplate tilt angle is given by $\theta (t=0)=26.55^o$ when the gel backbone is fully saturated with complexed copper ions. Then, using Eq. \ref{eq:height}, the initial unitless height of the hydrogel is $h(t=0)=0.9057$ and $u_z (z=1, t=0) = h(t=0)-1= -\gamma \phi^{(b)} = -0.0943\,.$ With, $\phi^{(b)}(z, t=0)=0.018$ and is uniform across the gel, which yields $\gamma=5.24\,.$ Since each COO$-$ group is occupied by one proton H$^+$ to form a stable COOH group in an acidic environment, $\lim_{t\rightarrow\infty}\phi^{(b)}_+(z, t)=\phi^\ast=0.036\,.$

The other chemical stress prefactor $\chi$ can be estimated based on the fact that the gel experimentally returns to its initial height when it reaches final equilibrium with all bound copper ions replaced by the protons at $t\rightarrow\infty\,.$ This observation can be understood by considering the binding distances of the COO$^--$Cu$^{2+}-$COO$^-$ chelates and the COOH groups: Even though Cu$^{2+}$ has a 63.5 times higher mass than H$^+$, the binding distance of a COO$^--$Cu$^{2+}-$COO$^-$ crosslink is not proportionally longer than the hydrogen bonds involved in the association of two neighboring COOH groups. For COOH$-$COOH hydrogen bonding, the center-to-center binding distance is approximately 1.6\AA~\cite{Tzeli2011}. For COO$^--$Cu$^{2+}$, the center-to-center binding distance is approximately 2\AA~\cite{Torres2011}. Although an exact comparison is difficult to make due to the bond angles, these numbers imply that the binding distances are quite similar for the two crosslinking types, leading to initial and final heights of the gel to be roughly equal. Then, the estimation $\lim_{t\rightarrow\infty}\phi^{(b)}_+(z, t)=\phi^\ast=0.036$ leads to $\chi=2.62$ per the boundary condition in Eq.~14 ($u_z(z)=z u_z(z=1)$ at equilibrium at $t=0$ and $t\rightarrow\infty$).

We also estimated the total cross-sectional bound copper in the gel $\phi^{(b)}_{total}$ by linearly interpolating the red channel value (r-value) from the optical microscopy of the gel samples with highest r-value corresponding to $\phi^{(b)}=0$ and lowest r-value corresponding to $\phi^{(b)}=0.018$ (Fig.~2b,~d).

\section{Validity of linear poroelastic model}

We here assess the validity of the linear elastic approximation to the gel deformations by comparing the linear strain terms with the components of the nonlinear deformation tensor underlying classical rubber elasticity (the Neo-Hookean solid model).  The nature of the theory we present in this work is fundamentally different than the theory we presented in Ref.~~\cite{Korevaar2020_2}, where we developed a thin-film theory for the ion-release induced deformation waves along PAA gel films. The theory was agnostic to the interface permeability and thus assumed a temporary osmotic imbalance of the ions based on van't Hoff's law. Here this assumption is relaxed and the interface permeability is explicitly taken into account by resolving the $z-$dependence of all the variables with the boundary conditions in Eqs.~10--12.

For a Neo-Hookean solid, the deformation tensor $\lambda_{ij}\equiv \partial R_i/\partial R_{0j}$ ($R_{0j}:$ $j-$component of the position vector of a material point in the reference configuration, $R_i:$ $i-$component of the position vector of the same material point after deformation, i.e., in the target configuration) can be expressed in terms of the $i-$component of the matrix displacement vector $u_i$ as ($\nabla_j\equiv \partial/\partial R_{0j}$, $\delta_{ij}:$ Kronecker delta)
\begin{equation}
    \label{eq:deformation_tensor}
    \lambda_{ij}=\delta_{ij}+\nabla_j u_i\,.
\end{equation}
The symmetric nonlinear elastic strain tensor is expressed in terms of the deformation tensor and the gradients of the displacement vector components, respectively, as
\begin{equation}
    \label{eq:strain_tensor}
    \epsilon_{ij}\equiv\frac{1}{2}\left(\lambda_{ik} \lambda_{kj}-\delta_{ij}\right) =\frac{1}{2}\left(\nabla_j u_i+ \nabla_i u_j+ \nabla_j u_i \nabla_i u_j\right)\,.
\end{equation}
%by the elastic strain tensor $\epsilon_{ij}\equiv \left(\nabla_j u_i+ \nabla_i u_j+ \nabla_j u_i \nabla_i u_j\right)/2$
For small displacements, Eq.~\ref{eq:strain_tensor} reduces to $\epsilon_{ij}\equiv \left(\nabla_j u_i+ \nabla_i u_j\right)/2$ in the linear limit as given before Eq.~2 in the main text. The linear approximation is valid when the inequalities $\nabla_j u_i\nabla_i u_j\ll \nabla_j u_i$ and $\nabla_j u_i\nabla_i u_j\ll \nabla_i u_j$ hold.

For uniaxial deformations shown in Fig.~2 (black curves), the time-dependent linear strain tensor has only one component, i.e., $\epsilon_{zz}=\partial u_z/\partial z\,.$ At a given time, The strain at a given time is always highest at the gel-supernatant interface (since it is a free boundary) and will be equal to $h(t)/H$ ($h(t):$ time-dependent gel height, $H:$ equilibrium gel height). In Fig.~2a, the maximum strain over the duration of the entire deformation cycle is $(\partial u_z/\partial z)_{max}=[h(\tau_{total})-h(0)]/h(0)=(0.985-0.905)/0.905\sim 9\%$ where $\tau_{total}\approx 0.82 \tau$ is defined as the total copper decomplexation time in the main text ($\tau:$ poroelastic deformation timescale, a.k.a. solvent absorption timescale). Then, the nonlinear correction to the strain is proportional to $(\partial u_z/\partial z)_{max}^2\sim 0.8\%$ at most. The linear poroelastic theory is therefore valid for a less than $1\%$ deviation in the strain and height profiles.

\section{Analytical solution for diffusive relaxation}

To determine the nature of the height decay in Fig.~1a from the maximum $h_{max}=0.985$ at $t=0.82\tau\,,$ here we probe a hypothetical purely diffusive relaxation dynamics by neglecting the second, third, and last terms on the right-hand side of Eq.~13. This is because the bound copper is instantaneously replaced by the acid on the gel backbone ($r\gg 1$) throughout swelling and subsequent relaxation. Then, the sum of the second and third terms is negligible, and the flux condition in Eq.~14 is roughly constant because of the rapid displacement of copper by acid, leaving the chemical stresses almost unaltered for the values of $\gamma\,,\chi$ calculated above and listed in Table~\ref{table:simulation_parameters}. Solving for purely diffusive relaxation allows us to classify whether the height decay is diffusive, subdiffusive, or superdiffusive. 

The analytical solution to the diffusion equation $\partial u_z/\partial t=\partial^2 u_z/\partial z^2$ subject to the boundary conditions in Eq.~14 leads to the solution for the gel height 

\begin{equation}
    h(t)=1+u_z(1, t)=1+c+\sum_{n=0}^{\infty}(-1)^n A_n e^{-k_n^2 t}\,,\quad k_n=\frac{(2n+1) \pi}{2}\,,\quad c\equiv- \gamma\phi^{(b)} - \chi\phi^{(b)}_+\approx \text{const.}\,,
    \label{eq:diffusion_solution}
\end{equation}
where the series coefficients $A_n$ can be found by using the initial condition $u_z=u_z(t=0.82\tau)$ obtained from the full uniaxial deformation model. Regardless, we are interested in the smallest timescale that gives the fastest decay as the leading order term of Eq.~\ref{eq:diffusion_solution}, which is 
\begin{equation}
\tau_D/\tau=k_0^{-2}=4/\pi^2\approx 0.4\,.    
\end{equation}
The second smallest timescale proportional to $k_1^{-2}=(2/3\pi)^2$ is nearly ten times smaller than $\tau_D\,,$ so we ignore the higher order terms in Eq.~\ref{eq:diffusion_solution}. Then, if the numerical solution of Eq.~13 yields a bigger (smaller) timescale than $\tau_D\,,$ it corresponds to a subdiffusive (superdiffusive) dynamics. Therefore, all decay timescales shown in Fig.~\ref{timescales} point to the subdiffusive relaxation.

\section{Gel deformation hysteresis}

The successive swelling and deswelling stages give rise to the hysteresis of the height-averaged gel pressure $p_{avg}$ with respect to the relative change in the gel height $h_r$ (Fig.~\ref{fig:results_hysteresis}). The area enclosed by a hysteresis curve $C$ determines the mechanical work per area done by the gel-supernatant system to its environment, defined as $\Delta w\equiv - (H \bar{p})^{-1} \int_C p_{avg} dh_r$ in unitless form. In our model, the sign of $\Delta w$ is governed by the competition between the energy gain due to the acid-driven release of copper, which interacts repulsively with the gel backbone in its free state, and the energy loss because of the viscous dissipation associated with the poroelastic deformations. The hysteresis curve for strong acid delivery is shown in Fig.~\ref{fig:results_hysteresis}a where we find $\Delta w=5.62\times 10^{-4}>0\,.$ The positive sign indicates that the energy lost to dissipation dominates the energy gain due to copper release, enabling the system to do net work to its environment. This amount of work yields a power density of $\sim 11.1$~mW/kg for the gel sample ($3.2$~mg dry weight over an area $L^2\,;$ see Table~\ref{table:simulation_parameters}) that completes its deformation cycle roughly over $4\tau\,.$ For comparison, other PAA microgels were shown to produce a power output of $260$ mW/kg as well as certain porous gels that exhibited $260$ mW/kg with 300\% strain and 0.2 s$^{-1}$ strain rates~\cite{Arens2017_2, Choudhary2022_2}. The main reason for the low power output in our case is that our experiments and theory are concerned with the linear elastic regime. On the other hand, the hysteresis curve for weak acid delivery, shown in Fig.~\ref{fig:results_hysteresis}b, yields $\Delta w=-5.15\times 10^{-4}<0$ ($\Delta W\approx -32\mu J$ over an area $L^2$) because of the negligible swelling as opposed to the considerable copper decomplexation (Fig.~1c,~d).

\begin{table*}
\caption{{\bf Simulation parameters}. The parameters denoted by ${(\ast)}$ correspond to the experimental values in Ref.~\cite{Korevaar2020_2}. For the parameters labeled by ${(\dagger)},$ the acid diffusivities were assumed $D_+=D^{(a)}_+\approx 3\times 10^{-8}m^2/s$ within the experimental range given in Refs.~\cite{Kinght2012_acid1, Agmon1995_acid2}, the copper diffusivities were taken from Ref.~\cite{Wu1990Copper} as $D_{Cu}\approx 6\times 10^{-10}m^2/s$ and $D^{(a)}_{Cu}\approx 10^{-9} m^2/s$ assuming that the Cu$^{2+}$ diffusion in the gel must be slower than in the supernatant due to the polymer presence. The diffusio-phoretic mobility is estimated as $D_{DP}=k_B T R_e^2/v_c\mu_f$ for steric repulsions between the polymer and the Cu$^{2+}$ ions with an exclusion radius $R_e=3\times 10^{-10}$~m, equivalent to $\eta_{DP}=R_e^2/k_f=0.225$ (Eq.~3)~\cite{Marbach2019_2}. The pressure scale $\bar{p}$ is estimated from Refs.~\cite{Baker2010, Sunyer2012, Denisin2016, Hossain2020}. The parameters denoted by ${(\ddagger)}$ are chosen to model an acid signal front as a step function in 1D, which is weakly perturbed with a Gaussian profile along the $x-$axis in 2D (Eq.~16). The value of $\Gamma^{(2\text{D})}$ approximates the built-in step function in Comsol per Eq.~15~\cite{Comsol}. The standard deviation $\lambda$ ensures that the width of the non-uniformity is comparable to the water capillary length $\ell_c\approx 10^{-3}$m. Among the parameters not labeled by a superscript, $k_f$ and $r$ are chosen in physically plausible ranges, whereas the low aspect ratio $\delta\ll 1$ models the gel as a thin film. The chemical stress moduli $\gamma$ and $\chi$ are found by fitting the experimental gel height of a copper-laden ($t=0$) and fully acid-complexed ($t\rightarrow\infty$) gel film, respectively (Sec.~\ref{post_processing}).}
\begin{center}
\small{\begin{tabular}{|c|c|c|}
\hline
\multicolumn{3}{|c|}{{\bf Parameters in real units}}\\   
   \hline
   gel height: $H=10^{-5}$m$~^{(\ast)}$ &  viscosity: $\mu_f=10^{-3}$Pa$\cdot$s$~^{(\ast)}$ & permeability: $k_f=4\times 10^{-19} m^2$ \\
   gel length: $L=2.5\times 10^{-2}$m$~^{(\ast)}$ & thermal energy: $k_B T\approx10^{-21}J$ & molecular volume: $v_c=10^{-29} m^3$$~^{(\ast)}$ \\ 
   \hline 
\end{tabular}}
\vskip 0.1in
\small{\begin{tabular}{|c|c|c|}
   \hline 
\multicolumn{3}{|c|}{{\bf Poroelastic deformation scales}}\\   
   \hline
   elastic constant: $\bar{p}\equiv 2\mu+\lambda=10^{5}$~Pa~$^{(\dagger)}$ & Timescale: $\tau\equiv \mu_f H^2/k_f \bar{p}=2.5 $ s & Diffusivity: $D\equiv H^2/\tau= 4\times 10^{-11} m^2/s$\\ \hline 
\end{tabular}}
\vskip 0.1in
\small{\begin{tabular}{|c|c|c|}   
   \hline
   \multicolumn{3}{|c|}{{\bf Unitless parameters for gel domain}}\\   
   \hline
   aspect ratio: $\delta\equiv H/L=4\times10^{-4}$ & reaction rate: $r\equiv \tilde{r}\tau=1.25\times 10^4$  & mobility: $\nu_{DP}\equiv D_{DP}/D=2.25\times10^2$$~^{(\dagger)}$\\ 
   Cu$^{2+}$ diffusivity: $\xi_{Cu}\equiv D_{Cu}/D=15$$~^{(\dagger)}$ & H$^{+}$ diffusivity: $\xi_{+}\equiv D_{+}/D=750$$~^{(\dagger)}$ & Cu$^{2+}$ stress modulus: $\gamma\equiv\tilde{\gamma}/\bar{p}=5.24$  \\
   Acid stress modulus: $\chi\equiv\tilde{\chi}/\bar{p}=2.62$ & polymer volume fraction: $\phi_p=0.04$$~^{(\ast)}$ & COO$-$ volume fraction: $\phi^\ast=0.036$$~^{(\ast)}$ \\ \hline
   \multicolumn{3}{|c|}{{\bf Unitless parameters for supernatant domain}}\\ \hline
    \multicolumn{2}{|c|}{Cu$^{2+}$ diffusivity: $\xi^{(a)}_{Cu}\equiv D^{(a)}_{Cu}/\alpha^2 D=4\times 10^{-5}$$~^{(\dagger)}$} & height ratio: $\alpha\equiv H^{(a)}/H=770$$~^{(\ast)}$\\ \multicolumn{2}{|c|}{H$^{+}$ diffusivity: $\xi^{(a)}_{+}\equiv D^{(a)}_{+}/\alpha^2 D=1.3\times 10^{-3}$$~^{(\dagger)}$} &\\ \hline
    \end{tabular}}
\vskip 0.1in
\small{\begin{tabular}{|c|c|c|}
    \hline
   \multicolumn{3}{|c|}{{\bf Unitless parameters for initial supernatant acid distribution $^{(\ddagger)}$}}\\ \hline
    \multicolumn{3}{|c|}{$\Gamma^{(1\text{D})}=1.25\times10^2\,,\Gamma^{(2\text{D})}=8\times10^3\,, z_0^{(1\text{D})}=1.04\,, h_1=1.04158\,, h_2=0.03158\,, \lambda=0.02\,.$}\\ \hline
   \end{tabular}}
\end{center}
\label{table:simulation_parameters}
\vspace{-10pt}
\end{table*}

\providecommand{\noopsort}[1]{}\providecommand{\singleletter}[1]{#1}%

\newpage

\begin{figure*}
 \includegraphics[width=1\textwidth]{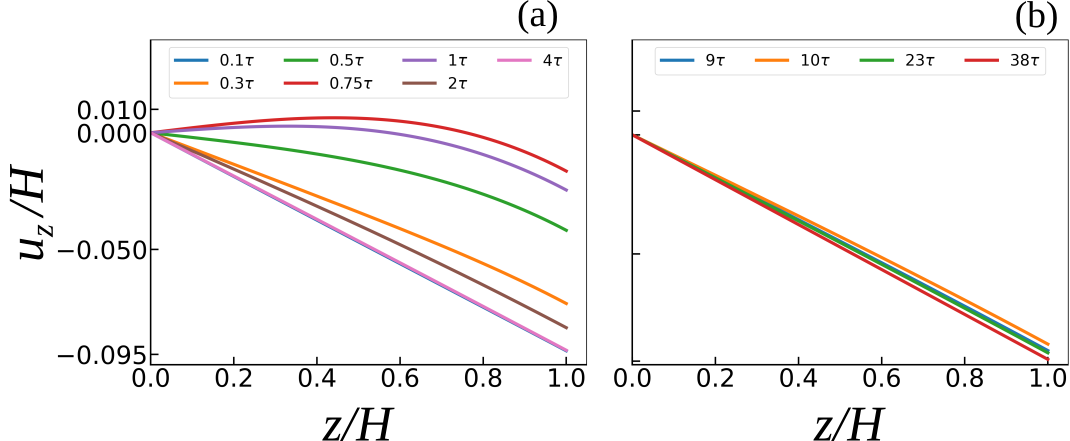}   
 \caption{\textbf{(a)}, \textbf{(b)} The evolution of the vertical displacement $u_z$ across the hydrogel for the addition of 1M and 0.05M acid, respectively.}
 \label{u_sppl}
\end{figure*}

\begin{figure*}
\includegraphics[width=1\textwidth]{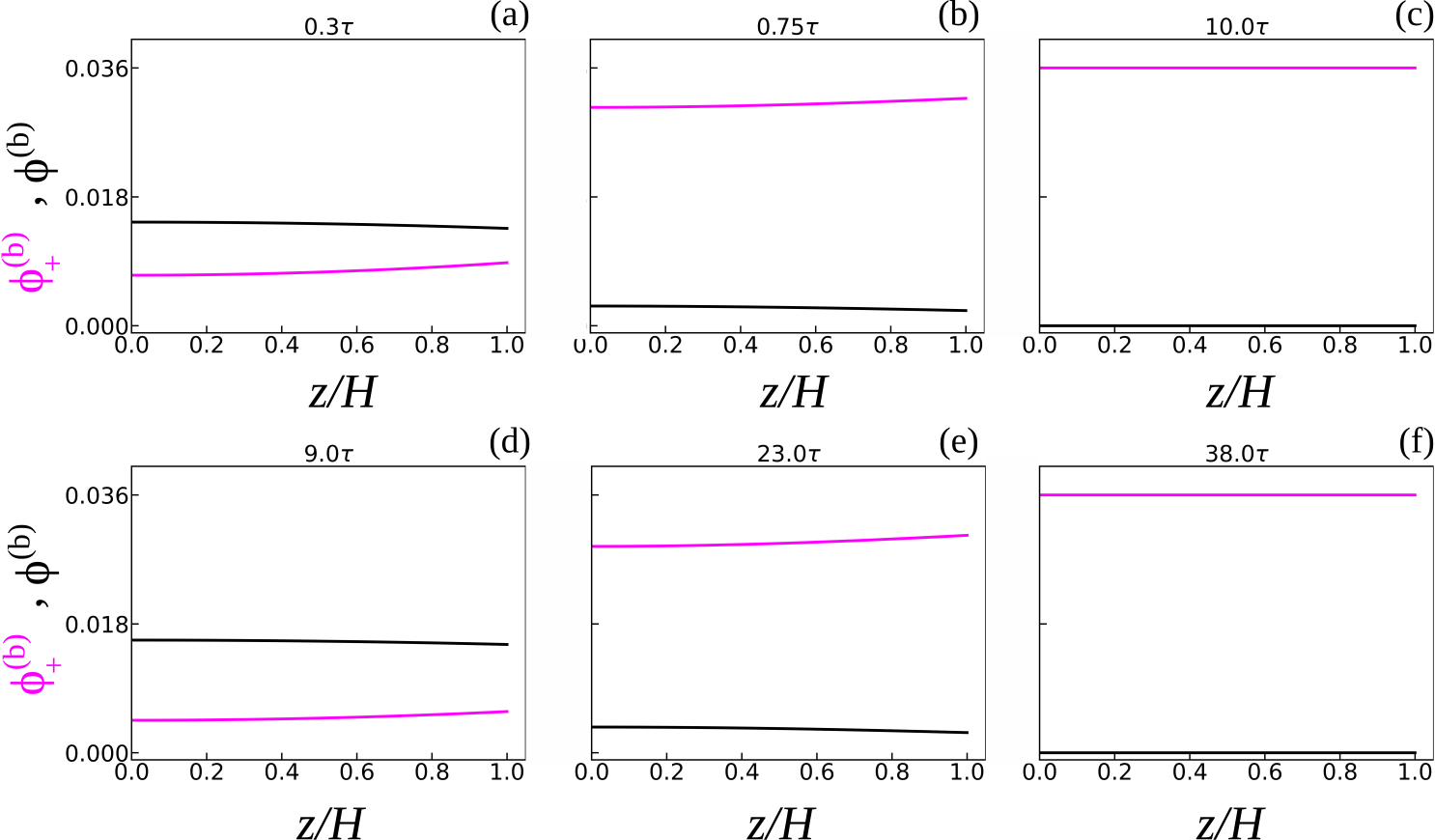}
\caption{\label{bound_ions_suppl} The time evolution of bound copper volume fraction $\phi^{(b)}$ (black curves) and bound acid volume fraction $\phi^{(b)}_+$ (magenta curves) across the hydrogel for the addition of \textbf{(a)}--\textbf{(c)} $1$M acid  and \textbf{(d)}--\textbf{(f)} $0.05$M acid. The addition step of $0.05$M acid starts at $t=8.4 \tau\,,$ following the addition of $0.01$M acid (see Fig.~2c, d).}
\end{figure*}

\begin{figure*}
 \includegraphics[width=1\textwidth]{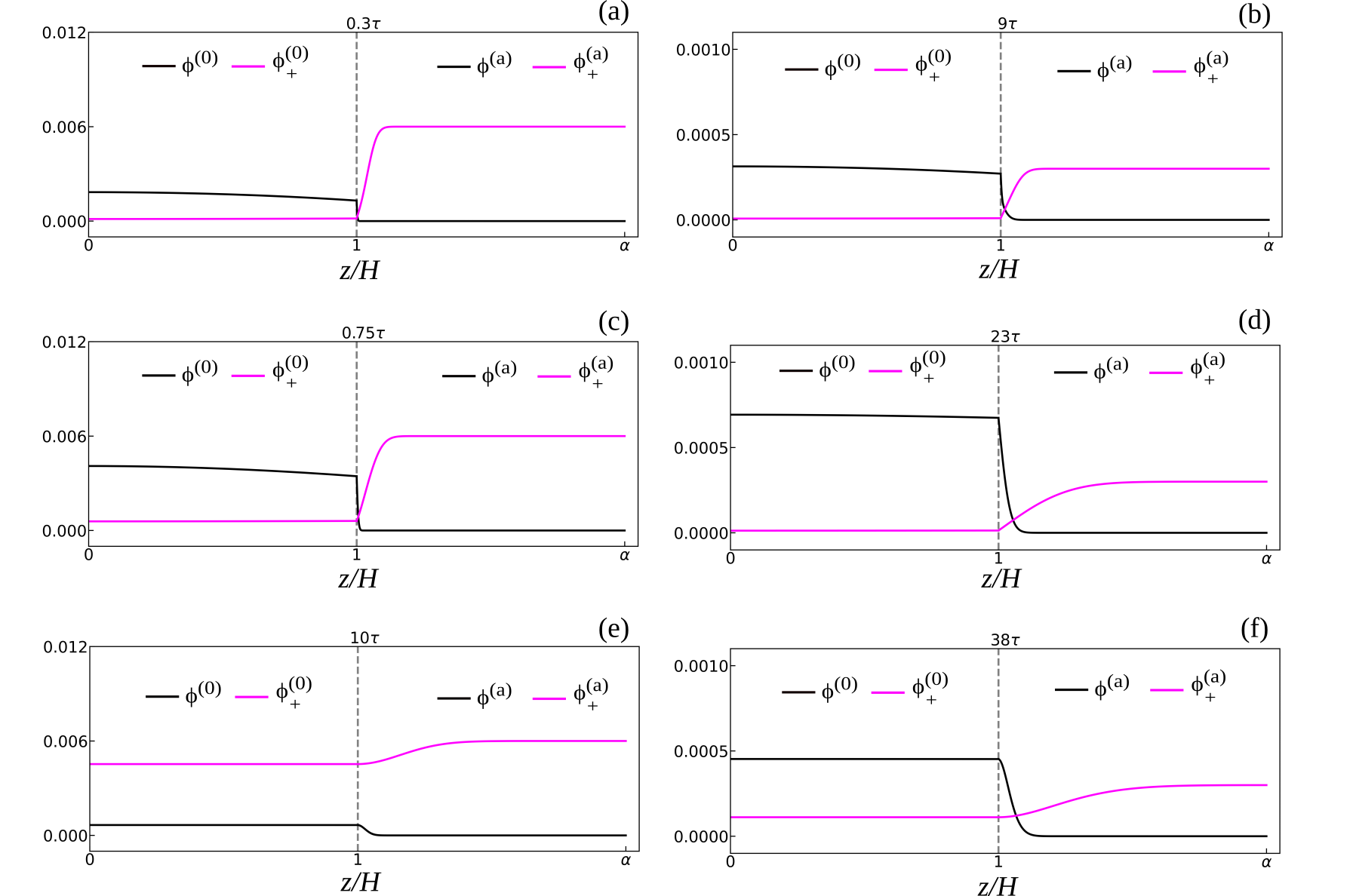} 
\caption{The time evolution of the free copper volume fraction (black curves) and the free acid volume fraction (magenta curves) across the hydrogel and supernatant domain for the addition of \textbf{(a)}, \textbf{(c)} and \textbf{(e)} $1$M acid and \textbf{(b)}, \textbf{(d)} and \textbf{(f)} $0.05$M acid. The addition step of $0.05$M acid starts at $t=8.4 \tau\,,$ following the addition of $0.01$M acid (see Fig.~2c, d). The dashed line at $z/H = 1$ indicates the gel-supernatant domain boundary.}
 \label{free_ions_suppl} 
\end{figure*}

\begin{figure*}
 \includegraphics[width=0.5\textwidth]{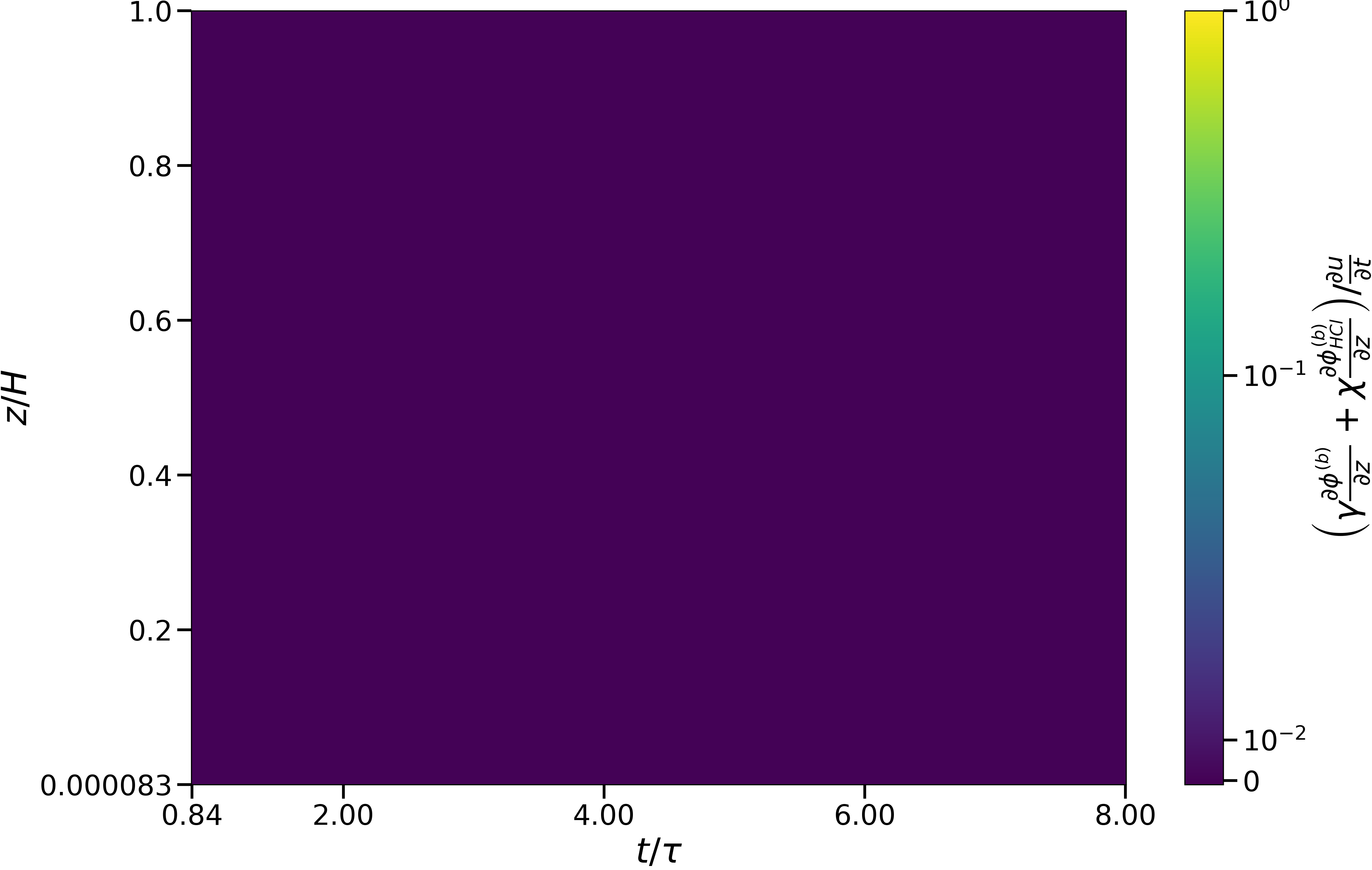} 
 \caption{The relative magnitude of the second and third terms on the right-hand side of Eq.~13 with respect to the time derivative of the vertical displacement.}
 \label{heatmap} 
\end{figure*}

\begin{figure*}
 \includegraphics[width=1\textwidth]{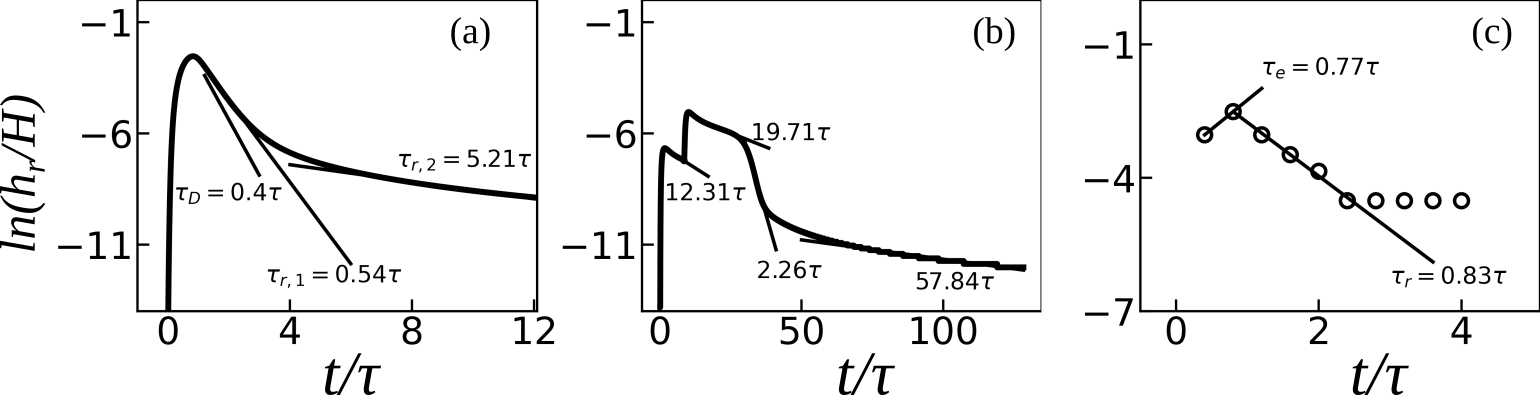} 
 \caption{The natural logarithm of the relative change of the unitless gel height $h_r(t)/H\equiv \left[h(t)-h(0)\right]/H$ with respect to the unitless time $t/\tau$ corresponding to the uniaxial deformation simulations for \textbf{(a)} strong acid (Fig.~2a), \textbf{(b)} weak acid(Fig.~2c), and \textbf{(c)} experimental swelling curve (Fig.~2a). The labels denote the timescales obtained by taking the inverse of the slope at each of the linear section. One exception is the diffusion timescale $\tau_D\,,$ which is shown as a guide to the eye to determine that $\tau_{r, 1}$ and $\tau_{r, 2}$ point to a subdiffusive dynamics. The experimental relaxation timescale $\tau_r$ in (c) is also subdiffusive as predicted by the theory.}
 \label{timescales} 
\end{figure*}

\begin{figure}
\centering
\includegraphics[width=0.5\textwidth]
{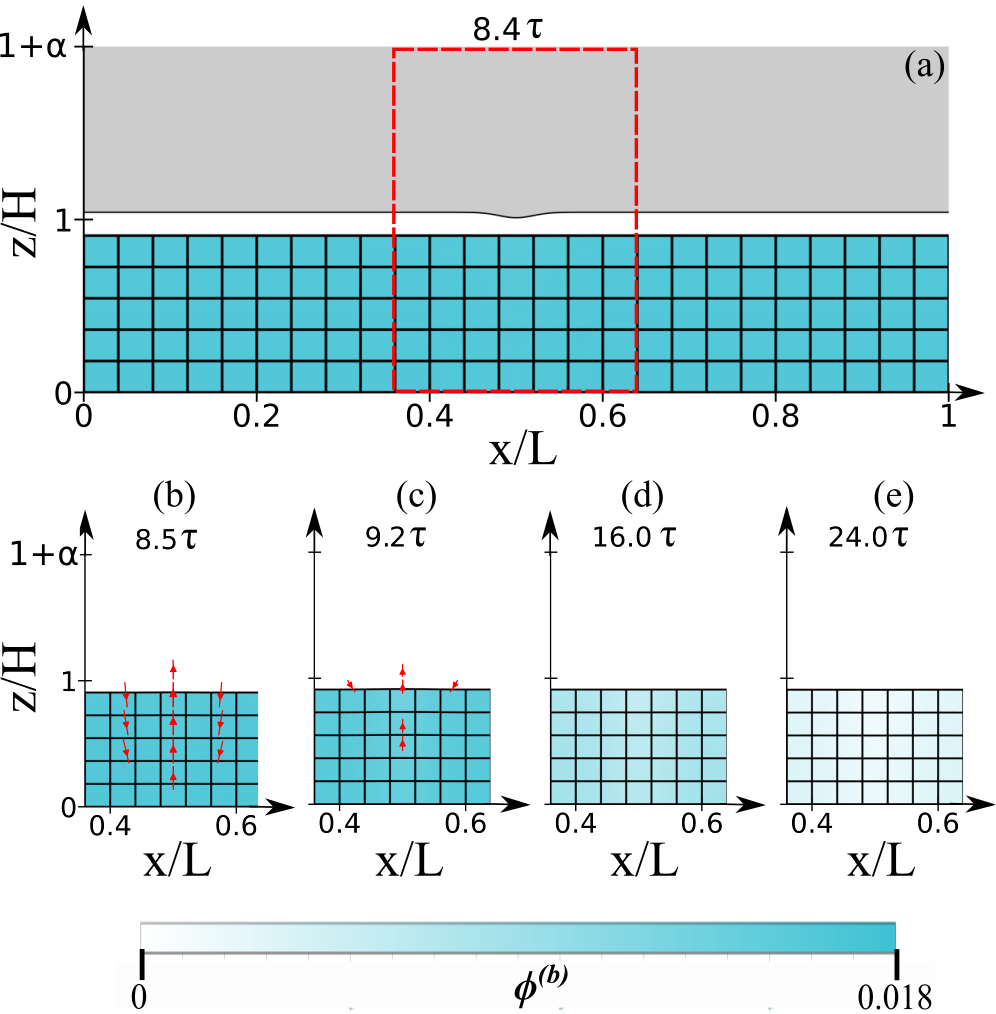}
\caption{\textbf{Gel response to a 0.05M 2D acid stimulus.}\textbf{(a)} Upon adding 0.05M acid with a perturbed front at $t=8.4\tau\,,$ the gel dynamics at \textbf{(b)} $t=8.5\tau$, \textbf{(c)} $t=9.2\tau$, \textbf{(d)} $t=16\tau$ and \textbf{(e)} $t=24\tau$ within the boxed region shown in (f). The red streamlines indicate the computed fluid flow in the lab frame (line width: logarithm of the flow speed, arrows: flow direction). The blue color scale indicates the volume fraction of the bound copper, and $\alpha=770$ (Table~\ref{table:simulation_parameters}).}
\label{fig:results2_weak}
\end{figure}

\begin{figure}[ht!]
\includegraphics[width=0.75\textwidth]{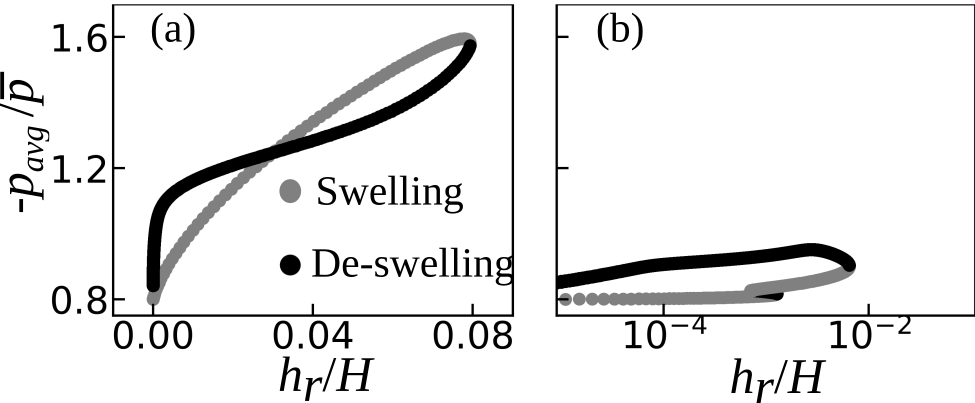}
\caption{\textbf{Gel deformation hysteresis.} The variation of the negative of the average gel pressure $p_{avg}/\bar{p}\equiv \int^1_0 p dz$ with respect to the relative change of the gel height $h_r(t)\equiv h(t)-h(0)$ in 1D (uniaxial deformations, Fig.~2a,~c) for \textbf{(a)} 1M acid delivery in the linear scale of and \textbf{(b)} successive delivery of 0.01M and 0.05M acid in the logarithmic scale of $h_r(t)\,.$}.
\label{fig:results_hysteresis}
\end{figure}

\begin{figure}[h!]
    \includegraphics[width=0.5\columnwidth]{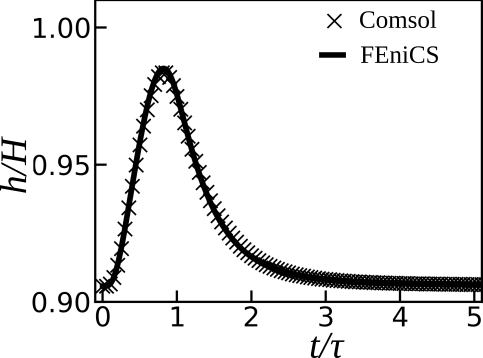}
    \caption{\label{library_comp} The simulated gel height $h(t)$ for uniaxial deformations by using the proprietary COMSOL Multiphysics 5.4 FEA Package~\cite{Comsol} and the open-source FEniCS FEA library on Python 3.9~\cite{AlnaesEtal2015}.}
    \label{fig:comparison}
\end{figure}


\begin{thebibliography}{50}%
\makeatletter
\providecommand \@ifxundefined [1]{%
 \@ifx{#1\undefined}
}%
\providecommand \@ifnum [1]{%
 \ifnum #1\expandafter \@firstoftwo
 \else \expandafter \@secondoftwo
 \fi
}%
\providecommand \@ifx [1]{%
 \ifx #1\expandafter \@firstoftwo
 \else \expandafter \@secondoftwo
 \fi
}%
\providecommand \natexlab [1]{#1}%
\providecommand \enquote  [1]{``#1''}%
\providecommand \bibnamefont  [1]{#1}%
\providecommand \bibfnamefont [1]{#1}%
\providecommand \citenamefont [1]{#1}%
\providecommand \href@noop [0]{\@secondoftwo}%
\providecommand \href [0]{\begingroup \@sanitize@url \@href}%
\providecommand \@href[1]{\@@startlink{#1}\@@href}%
\providecommand \@@href[1]{\endgroup#1\@@endlink}%
\providecommand \@sanitize@url [0]{\catcode `\\12\catcode `\$12\catcode `\&12\catcode `\#12\catcode `\^12\catcode `\_12\catcode `\%12\relax}%
\providecommand \@@startlink[1]{}%
\providecommand \@@endlink[0]{}%
\providecommand \url  [0]{\begingroup\@sanitize@url \@url }%
\providecommand \@url [1]{\endgroup\@href {#1}{\urlprefix }}%
\providecommand \urlprefix  [0]{URL }%
\providecommand \Eprint [0]{\href }%
\providecommand \doibase [0]{http://dx.doi.org/}%
\providecommand \selectlanguage [0]{\@gobble}%
\providecommand \bibinfo  [0]{\@secondoftwo}%
\providecommand \bibfield  [0]{\@secondoftwo}%
\providecommand \translation [1]{[#1]}%
\providecommand \BibitemOpen [0]{}%
\providecommand \bibitemStop [0]{}%
\providecommand \bibitemNoStop [0]{.\EOS\space}%
\providecommand \EOS [0]{\spacefactor3000\relax}%
\providecommand \BibitemShut  [1]{\csname bibitem#1\endcsname}%
\let\auto@bib@innerbib\@empty
%</preamble>
\bibitem [{\citenamefont {Kanahama}\ \emph {et~al.}(2023)\citenamefont {Kanahama}, \citenamefont {Tsugawa},\ and\ \citenamefont {Sato}}]{Kanahama2023}%
  \BibitemOpen
  \bibfield  {author} {\bibinfo {author} {\bibfnamefont {T.}~\bibnamefont {Kanahama}}, \bibinfo {author} {\bibfnamefont {S.}~\bibnamefont {Tsugawa}}, \ and\ \bibinfo {author} {\bibfnamefont {M.}~\bibnamefont {Sato}},\ }\href {\doibase 10.1038/s41598-023-29294-5} {\bibfield  {journal} {\bibinfo  {journal} {Scientific Reports}\ }\textbf {\bibinfo {volume} {13}},\ \bibinfo {pages} {2063} (\bibinfo {year} {2023})}\BibitemShut {NoStop}%
\bibitem [{\citenamefont {Schulgasser}\ and\ \citenamefont {Witztum}(1997)}]{Schulgasser1997}%
  \BibitemOpen
  \bibfield  {author} {\bibinfo {author} {\bibfnamefont {K.}~\bibnamefont {Schulgasser}}\ and\ \bibinfo {author} {\bibfnamefont {A.}~\bibnamefont {Witztum}},\ }\href {\doibase 10.1006/anbo.1997.0404} {\bibfield  {journal} {\bibinfo  {journal} {Annals of Botany}\ }\textbf {\bibinfo {volume} {80}},\ \bibinfo {pages} {35} (\bibinfo {year} {1997})}\BibitemShut {NoStop}%
\bibitem [{\citenamefont {Gottschalk}(1964)}]{GOTTSCHALK1964670}%
  \BibitemOpen
  \bibfield  {author} {\bibinfo {author} {\bibfnamefont {C.~W.}\ \bibnamefont {Gottschalk}},\ }\href {\doibase https://doi.org/10.1016/0002-9343(64)90179-2} {\bibfield  {journal} {\bibinfo  {journal} {The American Journal of Medicine}\ }\textbf {\bibinfo {volume} {36}},\ \bibinfo {pages} {670} (\bibinfo {year} {1964})},\ \bibinfo {note} {symposium on The Kidney}\BibitemShut {NoStop}%
\bibitem [{\citenamefont {Stephenson}(1972)}]{STEPHENSON197285}%
  \BibitemOpen
  \bibfield  {author} {\bibinfo {author} {\bibfnamefont {J.~L.}\ \bibnamefont {Stephenson}},\ }\href {\doibase https://doi.org/10.1038/ki.1972.75} {\bibfield  {journal} {\bibinfo  {journal} {Kidney International}\ }\textbf {\bibinfo {volume} {2}},\ \bibinfo {pages} {85} (\bibinfo {year} {1972})}\BibitemShut {NoStop}%
\bibitem [{\citenamefont {Brogioli}(2009)}]{Brogioli2009}%
  \BibitemOpen
  \bibfield  {author} {\bibinfo {author} {\bibfnamefont {D.}~\bibnamefont {Brogioli}},\ }\href {\doibase 10.1103/PhysRevLett.103.058501} {\bibfield  {journal} {\bibinfo  {journal} {Phys. Rev. Lett.}\ }\textbf {\bibinfo {volume} {103}},\ \bibinfo {pages} {058501} (\bibinfo {year} {2009})}\BibitemShut {NoStop}%
\bibitem [{\citenamefont {Tufa}\ \emph {et~al.}(2018)\citenamefont {Tufa}, \citenamefont {Pawlowski}, \citenamefont {Veerman}, \citenamefont {Bouzek}, \citenamefont {Fontananova}, \citenamefont {{di Profio}}, \citenamefont {Velizarov}, \citenamefont {{GoulÃƒÂ£o Crespo}}, \citenamefont {Nijmeijer},\ and\ \citenamefont {Curcio}}]{TUFA2018290}%
  \BibitemOpen
  \bibfield  {author} {\bibinfo {author} {\bibfnamefont {R.~A.}\ \bibnamefont {Tufa}}, \bibinfo {author} {\bibfnamefont {S.}~\bibnamefont {Pawlowski}}, \bibinfo {author} {\bibfnamefont {J.}~\bibnamefont {Veerman}}, \bibinfo {author} {\bibfnamefont {K.}~\bibnamefont {Bouzek}}, \bibinfo {author} {\bibfnamefont {E.}~\bibnamefont {Fontananova}}, \bibinfo {author} {\bibfnamefont {G.}~\bibnamefont {{di Profio}}}, \bibinfo {author} {\bibfnamefont {S.}~\bibnamefont {Velizarov}}, \bibinfo {author} {\bibfnamefont {J.}~\bibnamefont {{GoulÃƒÂ£o Crespo}}}, \bibinfo {author} {\bibfnamefont {K.}~\bibnamefont {Nijmeijer}}, \ and\ \bibinfo {author} {\bibfnamefont {E.}~\bibnamefont {Curcio}},\ }\href {\doibase https://doi.org/10.1016/j.apenergy.2018.04.111} {\bibfield  {journal} {\bibinfo  {journal} {Applied Energy}\ }\textbf {\bibinfo {volume} {225}},\ \bibinfo {pages} {290} (\bibinfo {year} {2018})}\BibitemShut {NoStop}%
\bibitem [{\citenamefont {Geise}\ \emph {et~al.}(2011)\citenamefont {Geise}, \citenamefont {Park}, \citenamefont {Sagle}, \citenamefont {Freeman},\ and\ \citenamefont {McGrath}}]{GEISE2011130}%
  \BibitemOpen
  \bibfield  {author} {\bibinfo {author} {\bibfnamefont {G.~M.}\ \bibnamefont {Geise}}, \bibinfo {author} {\bibfnamefont {H.~B.}\ \bibnamefont {Park}}, \bibinfo {author} {\bibfnamefont {A.~C.}\ \bibnamefont {Sagle}}, \bibinfo {author} {\bibfnamefont {B.~D.}\ \bibnamefont {Freeman}}, \ and\ \bibinfo {author} {\bibfnamefont {J.~E.}\ \bibnamefont {McGrath}},\ }\href {\doibase https://doi.org/10.1016/j.memsci.2010.11.054} {\bibfield  {journal} {\bibinfo  {journal} {Journal of Membrane Science}\ }\textbf {\bibinfo {volume} {369}},\ \bibinfo {pages} {130} (\bibinfo {year} {2011})}\BibitemShut {NoStop}%
\bibitem [{\citenamefont {Werber}\ \emph {et~al.}(2016)\citenamefont {Werber}, \citenamefont {Osuji},\ and\ \citenamefont {Elimelech}}]{Werber2016}%
  \BibitemOpen
  \bibfield  {author} {\bibinfo {author} {\bibfnamefont {J.~R.}\ \bibnamefont {Werber}}, \bibinfo {author} {\bibfnamefont {C.~O.}\ \bibnamefont {Osuji}}, \ and\ \bibinfo {author} {\bibfnamefont {M.}~\bibnamefont {Elimelech}},\ }\href {\doibase 10.1038/natrevmats.2016.18} {\bibfield  {journal} {\bibinfo  {journal} {Nature Reviews Materials}\ }\textbf {\bibinfo {volume} {1}},\ \bibinfo {pages} {16018} (\bibinfo {year} {2016})}\BibitemShut {NoStop}%
\bibitem [{\citenamefont {Logan}\ and\ \citenamefont {Elimelech}(2012)}]{Logan2012}%
  \BibitemOpen
  \bibfield  {author} {\bibinfo {author} {\bibfnamefont {B.~E.}\ \bibnamefont {Logan}}\ and\ \bibinfo {author} {\bibfnamefont {M.}~\bibnamefont {Elimelech}},\ }\href {\doibase 10.1038/nature11477} {\bibfield  {journal} {\bibinfo  {journal} {Nature}\ }\textbf {\bibinfo {volume} {488}},\ \bibinfo {pages} {313} (\bibinfo {year} {2012})}\BibitemShut {NoStop}%
\bibitem [{\citenamefont {Siria}\ \emph {et~al.}(2013)\citenamefont {Siria}, \citenamefont {Poncharal}, \citenamefont {Biance}, \citenamefont {Fulcrand}, \citenamefont {Blase}, \citenamefont {Purcell},\ and\ \citenamefont {Bocquet}}]{Siria2013}%
  \BibitemOpen
  \bibfield  {author} {\bibinfo {author} {\bibfnamefont {A.}~\bibnamefont {Siria}}, \bibinfo {author} {\bibfnamefont {P.}~\bibnamefont {Poncharal}}, \bibinfo {author} {\bibfnamefont {A.-L.}\ \bibnamefont {Biance}}, \bibinfo {author} {\bibfnamefont {R.}~\bibnamefont {Fulcrand}}, \bibinfo {author} {\bibfnamefont {X.}~\bibnamefont {Blase}}, \bibinfo {author} {\bibfnamefont {S.~T.}\ \bibnamefont {Purcell}}, \ and\ \bibinfo {author} {\bibfnamefont {L.}~\bibnamefont {Bocquet}},\ }\href {\doibase 10.1038/nature11876} {\bibfield  {journal} {\bibinfo  {journal} {Nature}\ }\textbf {\bibinfo {volume} {494}},\ \bibinfo {pages} {455} (\bibinfo {year} {2013})}\BibitemShut {NoStop}%
\bibitem [{\citenamefont {Siria}\ \emph {et~al.}(2017)\citenamefont {Siria}, \citenamefont {Bocquet},\ and\ \citenamefont {Bocquet}}]{Siria2017}%
  \BibitemOpen
  \bibfield  {author} {\bibinfo {author} {\bibfnamefont {A.}~\bibnamefont {Siria}}, \bibinfo {author} {\bibfnamefont {M.-L.}\ \bibnamefont {Bocquet}}, \ and\ \bibinfo {author} {\bibfnamefont {L.}~\bibnamefont {Bocquet}},\ }\href {\doibase 10.1038/s41570-017-0091} {\bibfield  {journal} {\bibinfo  {journal} {Nature Reviews Chemistry}\ }\textbf {\bibinfo {volume} {1}},\ \bibinfo {pages} {0091} (\bibinfo {year} {2017})}\BibitemShut {NoStop}%
\bibitem [{\citenamefont {Kedem}\ and\ \citenamefont {Katchalsky}(1961)}]{Kedem1961}%
  \BibitemOpen
  \bibfield  {author} {\bibinfo {author} {\bibfnamefont {O.}~\bibnamefont {Kedem}}\ and\ \bibinfo {author} {\bibfnamefont {A.}~\bibnamefont {Katchalsky}},\ }\href {\doibase 10.1085/jgp.45.1.143} {\bibfield  {journal} {\bibinfo  {journal} {Journal of General Physiology}\ }\textbf {\bibinfo {volume} {45}},\ \bibinfo {pages} {143} (\bibinfo {year} {1961})}\BibitemShut {NoStop}%
\bibitem [{\citenamefont {Zhang}\ and\ \citenamefont {Khademhosseini}(2017)}]{Zhang2017}%
  \BibitemOpen
  \bibfield  {author} {\bibinfo {author} {\bibfnamefont {Y.~S.}\ \bibnamefont {Zhang}}\ and\ \bibinfo {author} {\bibfnamefont {A.}~\bibnamefont {Khademhosseini}},\ }\href {\doibase 10.1126/science.aaf3627} {\enquote {\bibinfo {title} {Advances in engineering hydrogels},}\ } (\bibinfo {year} {2017})\BibitemShut {NoStop}%
\bibitem [{\citenamefont {Ulijn}\ \emph {et~al.}(2007)\citenamefont {Ulijn}, \citenamefont {Bibi}, \citenamefont {Jayawarna}, \citenamefont {Thornton}, \citenamefont {Todd}, \citenamefont {Mart}, \citenamefont {Smith},\ and\ \citenamefont {Gough}}]{Ulijn2007}%
  \BibitemOpen
  \bibfield  {author} {\bibinfo {author} {\bibfnamefont {R.~V.}\ \bibnamefont {Ulijn}}, \bibinfo {author} {\bibfnamefont {N.}~\bibnamefont {Bibi}}, \bibinfo {author} {\bibfnamefont {V.}~\bibnamefont {Jayawarna}}, \bibinfo {author} {\bibfnamefont {P.~D.}\ \bibnamefont {Thornton}}, \bibinfo {author} {\bibfnamefont {S.~J.}\ \bibnamefont {Todd}}, \bibinfo {author} {\bibfnamefont {R.~J.}\ \bibnamefont {Mart}}, \bibinfo {author} {\bibfnamefont {A.~M.}\ \bibnamefont {Smith}}, \ and\ \bibinfo {author} {\bibfnamefont {J.~E.}\ \bibnamefont {Gough}},\ }\href@noop {} {\enquote {\bibinfo {title} {Bioresponsive hydrogels},}\ } (\bibinfo {year} {2007})\BibitemShut {NoStop}%
\bibitem [{\citenamefont {Shim}\ \emph {et~al.}(2012)\citenamefont {Shim}, \citenamefont {Kim}, \citenamefont {Heo}, \citenamefont {Jeon},\ and\ \citenamefont {Yang}}]{Shim2012}%
  \BibitemOpen
  \bibfield  {author} {\bibinfo {author} {\bibfnamefont {T.~S.}\ \bibnamefont {Shim}}, \bibinfo {author} {\bibfnamefont {S.-H.}\ \bibnamefont {Kim}}, \bibinfo {author} {\bibfnamefont {C.-J.}\ \bibnamefont {Heo}}, \bibinfo {author} {\bibfnamefont {H.~C.}\ \bibnamefont {Jeon}}, \ and\ \bibinfo {author} {\bibfnamefont {S.-M.}\ \bibnamefont {Yang}},\ }\href {\doibase 10.1002/ange.201106723} {\bibfield  {journal} {\bibinfo  {journal} {Angewandte Chemie}\ }\textbf {\bibinfo {volume} {124}},\ \bibinfo {pages} {1449} (\bibinfo {year} {2012})}\BibitemShut {NoStop}%
\bibitem [{\citenamefont {Bordbar-Khiabani}\ and\ \citenamefont {Gasik}(2022)}]{Khiabani2022}%
  \BibitemOpen
  \bibfield  {author} {\bibinfo {author} {\bibfnamefont {A.}~\bibnamefont {Bordbar-Khiabani}}\ and\ \bibinfo {author} {\bibfnamefont {M.}~\bibnamefont {Gasik}},\ }\href {\doibase 10.3390/ijms23073665} {\bibfield  {journal} {\bibinfo  {journal} {International Journal of Molecular Sciences}\ }\textbf {\bibinfo {volume} {23}},\ \bibinfo {pages} {3665} (\bibinfo {year} {2022})}\BibitemShut {NoStop}%
\bibitem [{\citenamefont {Yuk}\ \emph {et~al.}(2017)\citenamefont {Yuk}, \citenamefont {Lin}, \citenamefont {Ma}, \citenamefont {Takaffoli}, \citenamefont {Fang},\ and\ \citenamefont {Zhao}}]{Yuk2017}%
  \BibitemOpen
  \bibfield  {author} {\bibinfo {author} {\bibfnamefont {H.}~\bibnamefont {Yuk}}, \bibinfo {author} {\bibfnamefont {S.}~\bibnamefont {Lin}}, \bibinfo {author} {\bibfnamefont {C.}~\bibnamefont {Ma}}, \bibinfo {author} {\bibfnamefont {M.}~\bibnamefont {Takaffoli}}, \bibinfo {author} {\bibfnamefont {N.~X.}\ \bibnamefont {Fang}}, \ and\ \bibinfo {author} {\bibfnamefont {X.}~\bibnamefont {Zhao}},\ }\href {\doibase 10.1038/ncomms14230} {\bibfield  {journal} {\bibinfo  {journal} {Nature Communications}\ }\textbf {\bibinfo {volume} {8}} (\bibinfo {year} {2017}),\ 10.1038/ncomms14230}\BibitemShut {NoStop}%
\bibitem [{\citenamefont {Ma}\ \emph {et~al.}(2013)\citenamefont {Ma}, \citenamefont {Guo}, \citenamefont {Anderson},\ and\ \citenamefont {Langer}}]{Ma2013}%
  \BibitemOpen
  \bibfield  {author} {\bibinfo {author} {\bibfnamefont {M.}~\bibnamefont {Ma}}, \bibinfo {author} {\bibfnamefont {L.}~\bibnamefont {Guo}}, \bibinfo {author} {\bibfnamefont {D.~G.}\ \bibnamefont {Anderson}}, \ and\ \bibinfo {author} {\bibfnamefont {R.}~\bibnamefont {Langer}},\ }\href {\doibase 10.1126/science.1230262} {\bibfield  {journal} {\bibinfo  {journal} {Science}\ }\textbf {\bibinfo {volume} {339}},\ \bibinfo {pages} {182} (\bibinfo {year} {2013})}\BibitemShut {NoStop}%
\bibitem [{\citenamefont {Dong}\ \emph {et~al.}(2006)\citenamefont {Dong}, \citenamefont {Agarwal}, \citenamefont {Beebe},\ and\ \citenamefont {Jiang}}]{Dong2006}%
  \BibitemOpen
  \bibfield  {author} {\bibinfo {author} {\bibfnamefont {L.}~\bibnamefont {Dong}}, \bibinfo {author} {\bibfnamefont {A.~K.}\ \bibnamefont {Agarwal}}, \bibinfo {author} {\bibfnamefont {D.~J.}\ \bibnamefont {Beebe}}, \ and\ \bibinfo {author} {\bibfnamefont {H.}~\bibnamefont {Jiang}},\ }\href {\doibase 10.1038/nature05024} {\bibfield  {journal} {\bibinfo  {journal} {Nature}\ }\textbf {\bibinfo {volume} {442}},\ \bibinfo {pages} {551} (\bibinfo {year} {2006})}\BibitemShut {NoStop}%
\bibitem [{\citenamefont {Zabow}\ \emph {et~al.}(2015)\citenamefont {Zabow}, \citenamefont {Dodd},\ and\ \citenamefont {Koretsky}}]{Zabow2015}%
  \BibitemOpen
  \bibfield  {author} {\bibinfo {author} {\bibfnamefont {G.}~\bibnamefont {Zabow}}, \bibinfo {author} {\bibfnamefont {S.~J.}\ \bibnamefont {Dodd}}, \ and\ \bibinfo {author} {\bibfnamefont {A.~P.}\ \bibnamefont {Koretsky}},\ }\href {\doibase 10.1038/nature14294} {\bibfield  {journal} {\bibinfo  {journal} {Nature}\ }\textbf {\bibinfo {volume} {520}},\ \bibinfo {pages} {73} (\bibinfo {year} {2015})}\BibitemShut {NoStop}%
\bibitem [{\citenamefont {Skoge}\ \emph {et~al.}(2014)\citenamefont {Skoge}, \citenamefont {Yue}, \citenamefont {Erickstad}, \citenamefont {Bae}, \citenamefont {Levine}, \citenamefont {Groisman}, \citenamefont {Loomis},\ and\ \citenamefont {Rappel}}]{Skoge2014}%
  \BibitemOpen
  \bibfield  {author} {\bibinfo {author} {\bibfnamefont {M.}~\bibnamefont {Skoge}}, \bibinfo {author} {\bibfnamefont {H.}~\bibnamefont {Yue}}, \bibinfo {author} {\bibfnamefont {M.}~\bibnamefont {Erickstad}}, \bibinfo {author} {\bibfnamefont {A.}~\bibnamefont {Bae}}, \bibinfo {author} {\bibfnamefont {H.}~\bibnamefont {Levine}}, \bibinfo {author} {\bibfnamefont {A.}~\bibnamefont {Groisman}}, \bibinfo {author} {\bibfnamefont {W.~F.}\ \bibnamefont {Loomis}}, \ and\ \bibinfo {author} {\bibfnamefont {W.-J.}\ \bibnamefont {Rappel}},\ }\href {\doibase 10.1073/pnas.1412197111} {\bibfield  {journal} {\bibinfo  {journal} {Proceedings of the National Academy of Sciences}\ }\textbf {\bibinfo {volume} {111}},\ \bibinfo {pages} {14448} (\bibinfo {year} {2014})}\BibitemShut {NoStop}%
\bibitem [{\citenamefont {Nerbonne}\ and\ \citenamefont {Kass}(2005)}]{Nerbonne2005}%
  \BibitemOpen
  \bibfield  {author} {\bibinfo {author} {\bibfnamefont {J.~M.}\ \bibnamefont {Nerbonne}}\ and\ \bibinfo {author} {\bibfnamefont {R.~S.}\ \bibnamefont {Kass}},\ }\href {\doibase 10.1152/physrev.00002.2005} {\bibfield  {journal} {\bibinfo  {journal} {Physiol Rev}\ }\textbf {\bibinfo {volume} {85}} (\bibinfo {year} {2005}),\ 10.1152/physrev.00002.2005}\BibitemShut {NoStop}%
\bibitem [{\citenamefont {PÃ©ter}\ \emph {et~al.}(2017)\citenamefont {P\'{e}ter}, \citenamefont {Glatz}, \citenamefont {Gudmann}, \citenamefont {Gombos}, \citenamefont {T\"{o}r\"{o}k}, \citenamefont {Horv\'{a}th}, \citenamefont {V\'{i}gh},\ and\ \citenamefont {Balogh}}]{Peter2017}%
  \BibitemOpen
  \bibfield  {author} {\bibinfo {author} {\bibfnamefont {M.}~\bibnamefont {P\'{e}ter}}, \bibinfo {author} {\bibfnamefont {A.}~\bibnamefont {Glatz}}, \bibinfo {author} {\bibfnamefont {P.}~\bibnamefont {Gudmann}}, \bibinfo {author} {\bibfnamefont {I.}~\bibnamefont {Gombos}}, \bibinfo {author} {\bibfnamefont {Z.}~\bibnamefont {T\"{o}r\"{o}k}}, \bibinfo {author} {\bibfnamefont {I.}~\bibnamefont {Horv\'{a}th}}, \bibinfo {author} {\bibfnamefont {L.}~\bibnamefont {V\'{i}gh}}, \ and\ \bibinfo {author} {\bibfnamefont {G.}~\bibnamefont {Balogh}},\ }\href {\doibase 10.1371/journal.pone.0173739} {\bibfield  {journal} {\bibinfo  {journal} {PLoS ONE}\ }\textbf {\bibinfo {volume} {12}} (\bibinfo {year} {2017}),\ 10.1371/journal.pone.0173739}\BibitemShut {NoStop}%
\bibitem [{\citenamefont {Choudhary}\ and\ \citenamefont {Raghavan}(2022)}]{Choudhary2022}%
  \BibitemOpen
  \bibfield  {author} {\bibinfo {author} {\bibfnamefont {H.}~\bibnamefont {Choudhary}}\ and\ \bibinfo {author} {\bibfnamefont {S.~R.}\ \bibnamefont {Raghavan}},\ }\href {\doibase 10.1021/acsami.2c00645} {\bibfield  {journal} {\bibinfo  {journal} {ACS Applied Materials \& Interfaces}\ }\textbf {\bibinfo {volume} {14}},\ \bibinfo {pages} {13733} (\bibinfo {year} {2022})},\ \bibinfo {note} {pMID: 35261243}\BibitemShut {NoStop}%
\bibitem [{\citenamefont {Zhang}\ \emph {et~al.}(2014)\citenamefont {Zhang}, \citenamefont {Zhou}, \citenamefont {Li}, \citenamefont {Sun}, \citenamefont {Kim}, \citenamefont {Fraden}, \citenamefont {Epstein},\ and\ \citenamefont {Xu}}]{epstein2014}%
  \BibitemOpen
  \bibfield  {author} {\bibinfo {author} {\bibfnamefont {Y.}~\bibnamefont {Zhang}}, \bibinfo {author} {\bibfnamefont {N.}~\bibnamefont {Zhou}}, \bibinfo {author} {\bibfnamefont {N.}~\bibnamefont {Li}}, \bibinfo {author} {\bibfnamefont {M.}~\bibnamefont {Sun}}, \bibinfo {author} {\bibfnamefont {D.}~\bibnamefont {Kim}}, \bibinfo {author} {\bibfnamefont {S.}~\bibnamefont {Fraden}}, \bibinfo {author} {\bibfnamefont {I.~R.}\ \bibnamefont {Epstein}}, \ and\ \bibinfo {author} {\bibfnamefont {B.}~\bibnamefont {Xu}},\ }\href {\doibase 10.1021/ja503665t} {\bibfield  {journal} {\bibinfo  {journal} {Journal of the American Chemical Society}\ }\textbf {\bibinfo {volume} {136}},\ \bibinfo {pages} {7341} (\bibinfo {year} {2014})},\ \bibinfo {note} {pMID: 24750134}\BibitemShut {NoStop}%
\bibitem [{\citenamefont {Arens}\ \emph {et~al.}(2017)\citenamefont {Arens}, \citenamefont {Wei\ss enfeld}, \citenamefont {Klein}, \citenamefont {Schlag},\ and\ \citenamefont {Wilhelm}}]{Arens2017}%
  \BibitemOpen
  \bibfield  {author} {\bibinfo {author} {\bibfnamefont {L.}~\bibnamefont {Arens}}, \bibinfo {author} {\bibfnamefont {F.}~\bibnamefont {Wei\ss enfeld}}, \bibinfo {author} {\bibfnamefont {C.~O.}\ \bibnamefont {Klein}}, \bibinfo {author} {\bibfnamefont {K.}~\bibnamefont {Schlag}}, \ and\ \bibinfo {author} {\bibfnamefont {M.}~\bibnamefont {Wilhelm}},\ }\href {\doibase https://doi.org/10.1002/advs.201700112} {\bibfield  {journal} {\bibinfo  {journal} {Advanced Science}\ }\textbf {\bibinfo {volume} {4}},\ \bibinfo {pages} {1700112} (\bibinfo {year} {2017})}\BibitemShut {NoStop}%
\bibitem [{\citenamefont {Korevaar}\ \emph {et~al.}(2020)\citenamefont {Korevaar}, \citenamefont {Kaplan}, \citenamefont {Grinthal}, \citenamefont {Rust},\ and\ \citenamefont {Aizenberg}}]{Korevaar2020}%
  \BibitemOpen
  \bibfield  {author} {\bibinfo {author} {\bibfnamefont {P.~A.}\ \bibnamefont {Korevaar}}, \bibinfo {author} {\bibfnamefont {C.~N.}\ \bibnamefont {Kaplan}}, \bibinfo {author} {\bibfnamefont {A.}~\bibnamefont {Grinthal}}, \bibinfo {author} {\bibfnamefont {R.~M.}\ \bibnamefont {Rust}}, \ and\ \bibinfo {author} {\bibfnamefont {J.}~\bibnamefont {Aizenberg}},\ }\href {\doibase 10.1038/s41467-019-14114-0} {\bibfield  {journal} {\bibinfo  {journal} {Nature Communications}\ }\textbf {\bibinfo {volume} {11}} (\bibinfo {year} {2020}),\ 10.1038/s41467-019-14114-0}\BibitemShut {NoStop}%
\bibitem [{\citenamefont {Velazquez}\ \emph {et~al.}(2011)\citenamefont {Velazquez}, \citenamefont {Peak-Chew}, \citenamefont {Fern\'{a}ndez}, \citenamefont {Neumann},\ and\ \citenamefont {Kay}}]{VELAZQUEZ20111252}%
  \BibitemOpen
  \bibfield  {author} {\bibinfo {author} {\bibfnamefont {F.}~\bibnamefont {Velazquez}}, \bibinfo {author} {\bibfnamefont {S.~Y.}\ \bibnamefont {Peak-Chew}}, \bibinfo {author} {\bibfnamefont {I.~S.}\ \bibnamefont {Fern\'{a}ndez}}, \bibinfo {author} {\bibfnamefont {C.~S.}\ \bibnamefont {Neumann}}, \ and\ \bibinfo {author} {\bibfnamefont {R.~R.}\ \bibnamefont {Kay}},\ }\href {\doibase https://doi.org/10.1016/j.chembiol.2011.08.003} {\bibfield  {journal} {\bibinfo  {journal} {Chemistry \& Biology}\ }\textbf {\bibinfo {volume} {18}},\ \bibinfo {pages} {1252} (\bibinfo {year} {2011})}\BibitemShut {NoStop}%
\bibitem [{\citenamefont {Mahadevan}\ and\ \citenamefont {Matsudaira}(2000)}]{Mahadevan2000}%
  \BibitemOpen
  \bibfield  {author} {\bibinfo {author} {\bibfnamefont {L.}~\bibnamefont {Mahadevan}}\ and\ \bibinfo {author} {\bibfnamefont {P.}~\bibnamefont {Matsudaira}},\ }\href@noop {} {\bibfield  {journal} {\bibinfo  {journal} {Science}\ }\textbf {\bibinfo {volume} {288}},\ \bibinfo {pages} {95} (\bibinfo {year} {2000})}\BibitemShut {NoStop}%
\bibitem [{\citenamefont {Palleau}\ \emph {et~al.}(2013)\citenamefont {Palleau}, \citenamefont {Morales}, \citenamefont {Dickey},\ and\ \citenamefont {Velev}}]{Palleau2013}%
  \BibitemOpen
  \bibfield  {author} {\bibinfo {author} {\bibfnamefont {E.}~\bibnamefont {Palleau}}, \bibinfo {author} {\bibfnamefont {D.}~\bibnamefont {Morales}}, \bibinfo {author} {\bibfnamefont {M.~D.}\ \bibnamefont {Dickey}}, \ and\ \bibinfo {author} {\bibfnamefont {O.~D.}\ \bibnamefont {Velev}},\ }\href {\doibase 10.1038/ncomms3257} {\bibfield  {journal} {\bibinfo  {journal} {Nature Communications}\ }\textbf {\bibinfo {volume} {4}},\ \bibinfo {pages} {2257} (\bibinfo {year} {2013})}\BibitemShut {NoStop}%
\bibitem [{\citenamefont {Longo}\ \emph {et~al.}(2014)\citenamefont {Longo}, \citenamefont {Olvera de~la Cruz},\ and\ \citenamefont {Szleifer}}]{Longo2014}%
  \BibitemOpen
  \bibfield  {author} {\bibinfo {author} {\bibfnamefont {G.~S.}\ \bibnamefont {Longo}}, \bibinfo {author} {\bibfnamefont {M.}~\bibnamefont {Olvera de~la Cruz}}, \ and\ \bibinfo {author} {\bibfnamefont {I.}~\bibnamefont {Szleifer}},\ }\href {\doibase 10.1063/1.4896562} {\bibfield  {journal} {\bibinfo  {journal} {The Journal of Chemical Physics}\ }\textbf {\bibinfo {volume} {141}},\ \bibinfo {pages} {124909} (\bibinfo {year} {2014})}\BibitemShut {NoStop}%
\bibitem [{\citenamefont {Longo}\ \emph {et~al.}(2016)\citenamefont {Longo}, \citenamefont {Olvera de~la Cruz},\ and\ \citenamefont {Szleifer}}]{Longo2016}%
  \BibitemOpen
  \bibfield  {author} {\bibinfo {author} {\bibfnamefont {G.~S.}\ \bibnamefont {Longo}}, \bibinfo {author} {\bibfnamefont {M.}~\bibnamefont {Olvera de~la Cruz}}, \ and\ \bibinfo {author} {\bibfnamefont {I.}~\bibnamefont {Szleifer}},\ }\href {\doibase 10.1039/C6SM01172A} {\bibfield  {journal} {\bibinfo  {journal} {Soft Matter}\ }\textbf {\bibinfo {volume} {12}},\ \bibinfo {pages} {8359} (\bibinfo {year} {2016})}\BibitemShut {NoStop}%
\bibitem [{\citenamefont {Drozdov}\ \emph {et~al.}(2016)\citenamefont {Drozdov}, \citenamefont {Sanporean},\ and\ \citenamefont {{deClaville Christiansen}}}]{Drozdov2016}%
  \BibitemOpen
  \bibfield  {author} {\bibinfo {author} {\bibfnamefont {A.}~\bibnamefont {Drozdov}}, \bibinfo {author} {\bibfnamefont {C.-G.}\ \bibnamefont {Sanporean}}, \ and\ \bibinfo {author} {\bibfnamefont {J.}~\bibnamefont {{deClaville Christiansen}}},\ }\href {\doibase https://doi.org/10.1016/j.mtcomm.2016.02.001} {\bibfield  {journal} {\bibinfo  {journal} {Materials Today Communications}\ }\textbf {\bibinfo {volume} {6}},\ \bibinfo {pages} {92} (\bibinfo {year} {2016})}\BibitemShut {NoStop}%
\bibitem [{si_()}]{si_dimensionless_eq}%
  \BibitemOpen
  \href@noop {} {}\bibinfo {note} {See the supplementary information for the supporting text, table for the physical and simulation parameters, figures, and movies.}\BibitemShut {Stop}%
\bibitem [{\citenamefont {Derjaguin}\ \emph {et~al.}(1993)\citenamefont {Derjaguin}, \citenamefont {Sidorenkov}, \citenamefont {Zubashchenko},\ and\ \citenamefont {Kiseleva}}]{DERJAGUIN1993138}%
  \BibitemOpen
  \bibfield  {author} {\bibinfo {author} {\bibfnamefont {B.}~\bibnamefont {Derjaguin}}, \bibinfo {author} {\bibfnamefont {G.}~\bibnamefont {Sidorenkov}}, \bibinfo {author} {\bibfnamefont {E.}~\bibnamefont {Zubashchenko}}, \ and\ \bibinfo {author} {\bibfnamefont {E.}~\bibnamefont {Kiseleva}},\ }\href {\doibase https://doi.org/10.1016/0079-6816(93)90023-O} {\bibfield  {journal} {\bibinfo  {journal} {Progress in Surface Science}\ }\textbf {\bibinfo {volume} {43}},\ \bibinfo {pages} {138} (\bibinfo {year} {1993})}\BibitemShut {NoStop}%
\bibitem [{\citenamefont {Marbach}\ and\ \citenamefont {Bocquet}(2019)}]{Marbach2019}%
  \BibitemOpen
  \bibfield  {author} {\bibinfo {author} {\bibfnamefont {S.}~\bibnamefont {Marbach}}\ and\ \bibinfo {author} {\bibfnamefont {L.}~\bibnamefont {Bocquet}},\ }\href {\doibase 10.1039/C8CS00420J} {\bibfield  {journal} {\bibinfo  {journal} {Chem. Soc. Rev.}\ }\textbf {\bibinfo {volume} {48}},\ \bibinfo {pages} {3102} (\bibinfo {year} {2019})}\BibitemShut {NoStop}%
\bibitem [{\citenamefont {Ram\'{i}rez-Hinestrosa}\ \emph {et~al.}(2020)\citenamefont {Ram\'{i}rez-Hinestrosa}, \citenamefont {Yoshida}, \citenamefont {Bocquet},\ and\ \citenamefont {Frenkel}}]{Hinestrosa2020}%
  \BibitemOpen
  \bibfield  {author} {\bibinfo {author} {\bibfnamefont {S.}~\bibnamefont {Ram\'{i}rez-Hinestrosa}}, \bibinfo {author} {\bibfnamefont {H.}~\bibnamefont {Yoshida}}, \bibinfo {author} {\bibfnamefont {L.}~\bibnamefont {Bocquet}}, \ and\ \bibinfo {author} {\bibfnamefont {D.}~\bibnamefont {Frenkel}},\ }\href {\doibase 10.1063/5.0007235} {\bibfield  {journal} {\bibinfo  {journal} {The Journal of Chemical Physics}\ }\textbf {\bibinfo {volume} {152}},\ \bibinfo {pages} {164901} (\bibinfo {year} {2020})},\ \Eprint {http://arxiv.org/abs/https://pubs.aip.org/aip/jcp/article-pdf/doi/10.1063/5.0007235/13421694/164901\_1\_online.pdf} {https://pubs.aip.org/aip/jcp/article-pdf/doi/10.1063/5.0007235/13421694/164901\_1\_online.pdf} \BibitemShut {NoStop}%
\bibitem [{\citenamefont {Banerjee}\ \emph {et~al.}(2016)\citenamefont {Banerjee}, \citenamefont {Williams}, \citenamefont {Azevedo}, \citenamefont {Helgeson},\ and\ \citenamefont {Squires}}]{Banerjee2016}%
  \BibitemOpen
  \bibfield  {author} {\bibinfo {author} {\bibfnamefont {A.}~\bibnamefont {Banerjee}}, \bibinfo {author} {\bibfnamefont {I.}~\bibnamefont {Williams}}, \bibinfo {author} {\bibfnamefont {R.~N.}\ \bibnamefont {Azevedo}}, \bibinfo {author} {\bibfnamefont {M.~E.}\ \bibnamefont {Helgeson}}, \ and\ \bibinfo {author} {\bibfnamefont {T.~M.}\ \bibnamefont {Squires}},\ }\href {\doibase 10.1073/pnas.1604743113} {\bibfield  {journal} {\bibinfo  {journal} {Proceedings of the National Academy of Sciences}\ }\textbf {\bibinfo {volume} {113}},\ \bibinfo {pages} {8612} (\bibinfo {year} {2016})},\ \Eprint {http://arxiv.org/abs/https://www.pnas.org/doi/pdf/10.1073/pnas.1604743113} {https://www.pnas.org/doi/pdf/10.1073/pnas.1604743113} \BibitemShut {NoStop}%
\bibitem [{\citenamefont {Banerjee}\ \emph {et~al.}(2020)\citenamefont {Banerjee}, \citenamefont {Tan},\ and\ \citenamefont {Squires}}]{Banerjee2020}%
  \BibitemOpen
  \bibfield  {author} {\bibinfo {author} {\bibfnamefont {A.}~\bibnamefont {Banerjee}}, \bibinfo {author} {\bibfnamefont {H.}~\bibnamefont {Tan}}, \ and\ \bibinfo {author} {\bibfnamefont {T.~M.}\ \bibnamefont {Squires}},\ }\href {\doibase 10.1103/PhysRevFluids.5.073701} {\bibfield  {journal} {\bibinfo  {journal} {Phys. Rev. Fluids}\ }\textbf {\bibinfo {volume} {5}},\ \bibinfo {pages} {073701} (\bibinfo {year} {2020})}\BibitemShut {NoStop}%
\bibitem [{\citenamefont {Sleeboom}\ \emph {et~al.}(2017)\citenamefont {Sleeboom}, \citenamefont {Voudouris}, \citenamefont {Punter}, \citenamefont {Aangenendt}, \citenamefont {Florea}, \citenamefont {van~der Schoot},\ and\ \citenamefont {Wyss}}]{Sleeboom2017}%
  \BibitemOpen
  \bibfield  {author} {\bibinfo {author} {\bibfnamefont {J.~J.~F.}\ \bibnamefont {Sleeboom}}, \bibinfo {author} {\bibfnamefont {P.}~\bibnamefont {Voudouris}}, \bibinfo {author} {\bibfnamefont {M.~T. J. J.~M.}\ \bibnamefont {Punter}}, \bibinfo {author} {\bibfnamefont {F.~J.}\ \bibnamefont {Aangenendt}}, \bibinfo {author} {\bibfnamefont {D.}~\bibnamefont {Florea}}, \bibinfo {author} {\bibfnamefont {P.}~\bibnamefont {van~der Schoot}}, \ and\ \bibinfo {author} {\bibfnamefont {H.~M.}\ \bibnamefont {Wyss}},\ }\href {\doibase 10.1103/PhysRevLett.119.098001} {\bibfield  {journal} {\bibinfo  {journal} {Phys. Rev. Lett.}\ }\textbf {\bibinfo {volume} {119}},\ \bibinfo {pages} {098001} (\bibinfo {year} {2017})}\BibitemShut {NoStop}%
\bibitem [{\citenamefont {Aangenendt}\ \emph {et~al.}(2020)\citenamefont {Aangenendt}, \citenamefont {Punter}, \citenamefont {Mulder}, \citenamefont {van~der Schoot},\ and\ \citenamefont {Wyss}}]{Aangenendt2020}%
  \BibitemOpen
  \bibfield  {author} {\bibinfo {author} {\bibfnamefont {F.~J.}\ \bibnamefont {Aangenendt}}, \bibinfo {author} {\bibfnamefont {M.~T. J. J.~M.}\ \bibnamefont {Punter}}, \bibinfo {author} {\bibfnamefont {B.~M.}\ \bibnamefont {Mulder}}, \bibinfo {author} {\bibfnamefont {P.}~\bibnamefont {van~der Schoot}}, \ and\ \bibinfo {author} {\bibfnamefont {H.~M.}\ \bibnamefont {Wyss}},\ }\href {\doibase 10.1103/PhysRevE.102.062606} {\bibfield  {journal} {\bibinfo  {journal} {Phys. Rev. E}\ }\textbf {\bibinfo {volume} {102}},\ \bibinfo {pages} {062606} (\bibinfo {year} {2020})}\BibitemShut {NoStop}%
\bibitem [{\citenamefont {Punter}\ \emph {et~al.}(2020)\citenamefont {Punter}, \citenamefont {Wyss},\ and\ \citenamefont {Mulder}}]{Punter2020}%
  \BibitemOpen
  \bibfield  {author} {\bibinfo {author} {\bibfnamefont {M.~T. J. J.~M.}\ \bibnamefont {Punter}}, \bibinfo {author} {\bibfnamefont {H.~M.}\ \bibnamefont {Wyss}}, \ and\ \bibinfo {author} {\bibfnamefont {B.~M.}\ \bibnamefont {Mulder}},\ }\href {\doibase 10.1103/PhysRevE.102.062607} {\bibfield  {journal} {\bibinfo  {journal} {Phys. Rev. E}\ }\textbf {\bibinfo {volume} {102}},\ \bibinfo {pages} {062607} (\bibinfo {year} {2020})}\BibitemShut {NoStop}%
\bibitem [{\citenamefont {Tong}\ and\ \citenamefont {Anderson}(1996)}]{Tong1996}%
  \BibitemOpen
  \bibfield  {author} {\bibinfo {author} {\bibfnamefont {J.}~\bibnamefont {Tong}}\ and\ \bibinfo {author} {\bibfnamefont {J.}~\bibnamefont {Anderson}},\ }\href {\doibase https://doi.org/10.1016/S0006-3495(96)79712-6} {\bibfield  {journal} {\bibinfo  {journal} {Biophysical Journal}\ }\textbf {\bibinfo {volume} {70}},\ \bibinfo {pages} {1505} (\bibinfo {year} {1996})}\BibitemShut {NoStop}%
\bibitem [{\citenamefont {Sachs}\ \emph {et~al.}(1994)\citenamefont {Sachs}, \citenamefont {Glucksberg}, \citenamefont {Jensen},\ and\ \citenamefont {Grotberg}}]{Sachs1994}%
  \BibitemOpen
  \bibfield  {author} {\bibinfo {author} {\bibfnamefont {J.~R.}\ \bibnamefont {Sachs}}, \bibinfo {author} {\bibfnamefont {M.~R.}\ \bibnamefont {Glucksberg}}, \bibinfo {author} {\bibfnamefont {O.~E.}\ \bibnamefont {Jensen}}, \ and\ \bibinfo {author} {\bibfnamefont {J.~B.}\ \bibnamefont {Grotberg}},\ }\href {\doibase 10.1115/1.2901525} {\bibfield  {journal} {\bibinfo  {journal} {Journal of Applied Mechanics}\ }\textbf {\bibinfo {volume} {61}},\ \bibinfo {pages} {726} (\bibinfo {year} {1994})}\BibitemShut {NoStop}%
\bibitem [{\citenamefont {Biot}(1941)}]{Biot1941}%
  \BibitemOpen
  \bibfield  {author} {\bibinfo {author} {\bibfnamefont {M.~A.}\ \bibnamefont {Biot}},\ }\href {\doibase 10.1063/1.1712886} {\bibfield  {journal} {\bibinfo  {journal} {Journal of Applied Physics}\ }\textbf {\bibinfo {volume} {12}},\ \bibinfo {pages} {155} (\bibinfo {year} {1941})}\BibitemShut {NoStop}%
\bibitem [{\citenamefont {Doi}(2013)}]{DoiSoftMatter}%
  \BibitemOpen
  \bibfield  {author} {\bibinfo {author} {\bibfnamefont {M.}~\bibnamefont {Doi}},\ }\href {\doibase 10.1093/acprof:oso/9780199652952.001.0001} {\emph {\bibinfo {title} {{Soft Matter Physics}}}}\ (\bibinfo  {publisher} {Oxford University Press},\ \bibinfo {year} {2013})\BibitemShut {NoStop}%
\bibitem [{\citenamefont {Hong}\ \emph {et~al.}(2008)\citenamefont {Hong}, \citenamefont {Zhao}, \citenamefont {Zhou},\ and\ \citenamefont {Suo}}]{hong2008}%
  \BibitemOpen
  \bibfield  {author} {\bibinfo {author} {\bibfnamefont {W.}~\bibnamefont {Hong}}, \bibinfo {author} {\bibfnamefont {X.}~\bibnamefont {Zhao}}, \bibinfo {author} {\bibfnamefont {J.}~\bibnamefont {Zhou}}, \ and\ \bibinfo {author} {\bibfnamefont {Z.}~\bibnamefont {Suo}},\ }\href {\doibase https://doi.org/10.1016/j.jmps.2007.11.010} {\bibfield  {journal} {\bibinfo  {journal} {Journal of the Mechanics and Physics of Solids}\ }\textbf {\bibinfo {volume} {56}},\ \bibinfo {pages} {1779} (\bibinfo {year} {2008})}\BibitemShut {NoStop}%
\bibitem [{\citenamefont {Hong}\ \emph {et~al.}(2009)\citenamefont {Hong}, \citenamefont {Liu},\ and\ \citenamefont {Suo}}]{hong2009}%
  \BibitemOpen
  \bibfield  {author} {\bibinfo {author} {\bibfnamefont {W.}~\bibnamefont {Hong}}, \bibinfo {author} {\bibfnamefont {Z.}~\bibnamefont {Liu}}, \ and\ \bibinfo {author} {\bibfnamefont {Z.}~\bibnamefont {Suo}},\ }\href {\doibase https://doi.org/10.1016/j.ijsolstr.2009.04.022} {\bibfield  {journal} {\bibinfo  {journal} {International Journal of Solids and Structures}\ }\textbf {\bibinfo {volume} {46}},\ \bibinfo {pages} {3282} (\bibinfo {year} {2009})}\BibitemShut {NoStop}%
\bibitem [{\citenamefont {MacMinn}\ \emph {et~al.}(2016)\citenamefont {MacMinn}, \citenamefont {Dufresne},\ and\ \citenamefont {Wettlaufer}}]{MacMinn2016}%
  \BibitemOpen
  \bibfield  {author} {\bibinfo {author} {\bibfnamefont {C.~W.}\ \bibnamefont {MacMinn}}, \bibinfo {author} {\bibfnamefont {E.~R.}\ \bibnamefont {Dufresne}}, \ and\ \bibinfo {author} {\bibfnamefont {J.~S.}\ \bibnamefont {Wettlaufer}},\ }\href {\doibase 10.1103/PhysRevApplied.5.044020} {\bibfield  {journal} {\bibinfo  {journal} {Phys. Rev. Appl.}\ }\textbf {\bibinfo {volume} {5}},\ \bibinfo {pages} {044020} (\bibinfo {year} {2016})}\BibitemShut {NoStop}%
\bibitem [{\citenamefont {Dayal}\ \emph {et~al.}(2013)\citenamefont {Dayal}, \citenamefont {Kuksenok},\ and\ \citenamefont {Balazs}}]{Balazs2013}%
  \BibitemOpen
  \bibfield  {author} {\bibinfo {author} {\bibfnamefont {P.}~\bibnamefont {Dayal}}, \bibinfo {author} {\bibfnamefont {O.}~\bibnamefont {Kuksenok}}, \ and\ \bibinfo {author} {\bibfnamefont {A.~C.}\ \bibnamefont {Balazs}},\ }\href {\doibase 10.1073/pnas.1213432110} {\bibfield  {journal} {\bibinfo  {journal} {Proceedings of the National Academy of Sciences}\ }\textbf {\bibinfo {volume} {110}},\ \bibinfo {pages} {431} (\bibinfo {year} {2013})}\BibitemShut {NoStop}%
\end{thebibliography}

\begin{thebibliography}{17}%
\makeatletter
\providecommand \@ifxundefined [1]{%
 \@ifx{#1\undefined}
}%
\providecommand \@ifnum [1]{%
 \ifnum #1\expandafter \@firstoftwo
 \else \expandafter \@secondoftwo
 \fi
}%
\providecommand \@ifx [1]{%
 \ifx #1\expandafter \@firstoftwo
 \else \expandafter \@secondoftwo
 \fi
}%
\providecommand \natexlab [1]{#1}%
\providecommand \enquote  [1]{``#1''}%
\providecommand \bibnamefont  [1]{#1}%
\providecommand \bibfnamefont [1]{#1}%
\providecommand \citenamefont [1]{#1}%
\providecommand \href@noop [0]{\@secondoftwo}%
\providecommand \href [0]{\begingroup \@sanitize@url \@href}%
\providecommand \@href[1]{\@@startlink{#1}\@@href}%
\providecommand \@@href[1]{\endgroup#1\@@endlink}%
\providecommand \@sanitize@url [0]{\catcode `\\12\catcode `\$12\catcode `\&12\catcode `\#12\catcode `\^12\catcode `\_12\catcode `\%12\relax}%
\providecommand \@@startlink[1]{}%
\providecommand \@@endlink[0]{}%
\providecommand \url  [0]{\begingroup\@sanitize@url \@url }%
\providecommand \@url [1]{\endgroup\@href {#1}{\urlprefix }}%
\providecommand \urlprefix  [0]{URL }%
\providecommand \Eprint [0]{\href }%
\providecommand \doibase [0]{https://doi.org/}%
\providecommand \selectlanguage [0]{\@gobble}%
\providecommand \bibinfo  [0]{\@secondoftwo}%
\providecommand \bibfield  [0]{\@secondoftwo}%
\providecommand \translation [1]{[#1]}%
\providecommand \BibitemOpen [0]{}%
\providecommand \bibitemStop [0]{}%
\providecommand \bibitemNoStop [0]{.\EOS\space}%
\providecommand \EOS [0]{\spacefactor3000\relax}%
\providecommand \BibitemShut  [1]{\csname bibitem#1\endcsname}%
\let\auto@bib@innerbib\@empty
%</preamble>
\bibitem [{\citenamefont {Flory}\ and\ \citenamefont {Rehner}(2004)}]{FloryRehner}%
  \BibitemOpen
  \bibfield  {author} {\bibinfo {author} {\bibfnamefont {P.~J.}\ \bibnamefont {Flory}}\ and\ \bibinfo {author} {\bibfnamefont {J.}~\bibnamefont {Rehner}, \bibfnamefont {John}},\ }\bibfield  {title} {\bibinfo {title} {{Statistical Mechanics of Cross‐Linked Polymer Networks II. Swelling}},\ }\href {https://doi.org/10.1063/1.1723792} {\bibfield  {journal} {\bibinfo  {journal} {The Journal of Chemical Physics}\ }\textbf {\bibinfo {volume} {11}},\ \bibinfo {pages} {521} (\bibinfo {year} {2004})},\ \Eprint {https://arxiv.org/abs/https://pubs.aip.org/aip/jcp/article-pdf/11/11/521/8106549/521\_1\_online.pdf} {https://pubs.aip.org/aip/jcp/article-pdf/11/11/521/8106549/521\_1\_online.pdf} \BibitemShut {NoStop}%
\bibitem [{\citenamefont {Korevaar}\ \emph {et~al.}(2020)\citenamefont {Korevaar}, \citenamefont {Kaplan}, \citenamefont {Grinthal}, \citenamefont {Rust},\ and\ \citenamefont {Aizenberg}}]{Korevaar2020_2}%
  \BibitemOpen
  \bibfield  {author} {\bibinfo {author} {\bibfnamefont {P.~A.}\ \bibnamefont {Korevaar}}, \bibinfo {author} {\bibfnamefont {C.~N.}\ \bibnamefont {Kaplan}}, \bibinfo {author} {\bibfnamefont {A.}~\bibnamefont {Grinthal}}, \bibinfo {author} {\bibfnamefont {R.~M.}\ \bibnamefont {Rust}},\ and\ \bibinfo {author} {\bibfnamefont {J.}~\bibnamefont {Aizenberg}},\ }\bibfield  {title} {\bibinfo {title} {Non-equilibrium signal integration in hydrogels},\ }\bibfield  {journal} {\bibinfo  {journal} {Nature Communications}\ }\textbf {\bibinfo {volume} {11}},\ \href {https://doi.org/10.1038/s41467-019-14114-0} {10.1038/s41467-019-14114-0} (\bibinfo {year} {2020})\BibitemShut {NoStop}%
\bibitem [{\citenamefont {Palleau}\ \emph {et~al.}(2013)\citenamefont {Palleau}, \citenamefont {Morales}, \citenamefont {Dickey},\ and\ \citenamefont {Velev}}]{Palleau2013_2}%
  \BibitemOpen
  \bibfield  {author} {\bibinfo {author} {\bibfnamefont {E.}~\bibnamefont {Palleau}}, \bibinfo {author} {\bibfnamefont {D.}~\bibnamefont {Morales}}, \bibinfo {author} {\bibfnamefont {M.~D.}\ \bibnamefont {Dickey}},\ and\ \bibinfo {author} {\bibfnamefont {O.~D.}\ \bibnamefont {Velev}},\ }\bibfield  {title} {\bibinfo {title} {Reversible patterning and actuation of hydrogels by electrically assisted ionoprinting},\ }\href {https://doi.org/10.1038/ncomms3257} {\bibfield  {journal} {\bibinfo  {journal} {Nature Communications}\ }\textbf {\bibinfo {volume} {4}},\ \bibinfo {pages} {2257} (\bibinfo {year} {2013})}\BibitemShut {NoStop}%
\bibitem [{\citenamefont {Alnaes}\ \emph {et~al.}(2015)\citenamefont {Alnaes}, \citenamefont {Blechta}, \citenamefont {Hake}, \citenamefont {Johansson}, \citenamefont {Kehlet}, \citenamefont {Logg}, \citenamefont {Richardson}, \citenamefont {Ring}, \citenamefont {Rognes},\ and\ \citenamefont {Wells}}]{AlnaesEtal2015}%
  \BibitemOpen
  \bibfield  {author} {\bibinfo {author} {\bibfnamefont {M.~S.}\ \bibnamefont {Alnaes}}, \bibinfo {author} {\bibfnamefont {J.}~\bibnamefont {Blechta}}, \bibinfo {author} {\bibfnamefont {J.}~\bibnamefont {Hake}}, \bibinfo {author} {\bibfnamefont {A.}~\bibnamefont {Johansson}}, \bibinfo {author} {\bibfnamefont {B.}~\bibnamefont {Kehlet}}, \bibinfo {author} {\bibfnamefont {A.}~\bibnamefont {Logg}}, \bibinfo {author} {\bibfnamefont {C.}~\bibnamefont {Richardson}}, \bibinfo {author} {\bibfnamefont {J.}~\bibnamefont {Ring}}, \bibinfo {author} {\bibfnamefont {M.~E.}\ \bibnamefont {Rognes}},\ and\ \bibinfo {author} {\bibfnamefont {G.~N.}\ \bibnamefont {Wells}},\ }\bibfield  {title} {\bibinfo {title} {The {FEniCS} project version 1.5},\ }\bibfield  {journal} {\bibinfo  {journal} {Archive of Numerical Software}\ }\textbf {\bibinfo {volume} {3}},\ \href {https://doi.org/10.11588/ans.2015.100.20553} {10.11588/ans.2015.100.20553} (\bibinfo {year} {2015})\BibitemShut {NoStop}%
\bibitem [{\citenamefont {{COMSOL AB, Stockholm, Sweden}}(2018)}]{Comsol}%
  \BibitemOpen
  \bibfield  {author} {\bibinfo {author} {\bibnamefont {{COMSOL AB, Stockholm, Sweden}}},\ }\href {https://www.comsol.com/} {\bibinfo {title} {{COMSOL Multiphysics® 5.4, www.comsol.com}}} (\bibinfo {year} {2018})\BibitemShut {NoStop}%
\bibitem [{\citenamefont {Tzeli}\ \emph {et~al.}(2011)\citenamefont {Tzeli}, \citenamefont {Theodorakopoulos}, \citenamefont {Petsalakis}, \citenamefont {Ajami},\ and\ \citenamefont {Rebek}}]{Tzeli2011}%
  \BibitemOpen
  \bibfield  {author} {\bibinfo {author} {\bibfnamefont {D.}~\bibnamefont {Tzeli}}, \bibinfo {author} {\bibfnamefont {G.}~\bibnamefont {Theodorakopoulos}}, \bibinfo {author} {\bibfnamefont {I.~D.}\ \bibnamefont {Petsalakis}}, \bibinfo {author} {\bibfnamefont {D.}~\bibnamefont {Ajami}},\ and\ \bibinfo {author} {\bibfnamefont {J.}~\bibnamefont {Rebek}},\ }\bibfield  {title} {\bibinfo {title} {Theoretical study of hydrogen bonding in homodimers and heterodimers of amide, boronic acid, and carboxylic acid, free and in encapsulation complexes},\ }\href {https://doi.org/10.1021/ja206555d} {\bibfield  {journal} {\bibinfo  {journal} {Journal of the American Chemical Society}\ }\textbf {\bibinfo {volume} {133}},\ \bibinfo {pages} {16977} (\bibinfo {year} {2011})},\ \bibinfo {note} {pMID: 21923158},\ \Eprint {https://arxiv.org/abs/https://doi.org/10.1021/ja206555d} {https://doi.org/10.1021/ja206555d} \BibitemShut {NoStop}%
\bibitem [{\citenamefont {Al\'{i}-Torres}\ \emph {et~al.}(2011)\citenamefont {Al\'{i}-Torres}, \citenamefont {Mar\'{e}chal}, \citenamefont {Rodr\'{i}guez-Santiago},\ and\ \citenamefont {Sodupe}}]{Torres2011}%
  \BibitemOpen
  \bibfield  {author} {\bibinfo {author} {\bibfnamefont {J.}~\bibnamefont {Al\'{i}-Torres}}, \bibinfo {author} {\bibfnamefont {J.-D.}\ \bibnamefont {Mar\'{e}chal}}, \bibinfo {author} {\bibfnamefont {L.}~\bibnamefont {Rodr\'{i}guez-Santiago}},\ and\ \bibinfo {author} {\bibfnamefont {M.}~\bibnamefont {Sodupe}},\ }\bibfield  {title} {\bibinfo {title} {Three dimensional models of {Cu$^{2+}$-A$\beta$(1-16)} complexes from computational approaches},\ }\href {https://doi.org/10.1021/ja203407v} {\bibfield  {journal} {\bibinfo  {journal} {Journal of the American Chemical Society}\ }\textbf {\bibinfo {volume} {133}},\ \bibinfo {pages} {15008} (\bibinfo {year} {2011})},\ \bibinfo {note} {pMID: 21846101},\ \Eprint {https://arxiv.org/abs/https://doi.org/10.1021/ja203407v} {https://doi.org/10.1021/ja203407v} \BibitemShut {NoStop}%
\bibitem [{\citenamefont {Arens}\ \emph {et~al.}(2017)\citenamefont {Arens}, \citenamefont {Weißenfeld}, \citenamefont {Klein}, \citenamefont {Schlag},\ and\ \citenamefont {Wilhelm}}]{Arens2017_2}%
  \BibitemOpen
  \bibfield  {author} {\bibinfo {author} {\bibfnamefont {L.}~\bibnamefont {Arens}}, \bibinfo {author} {\bibfnamefont {F.}~\bibnamefont {Weißenfeld}}, \bibinfo {author} {\bibfnamefont {C.~O.}\ \bibnamefont {Klein}}, \bibinfo {author} {\bibfnamefont {K.}~\bibnamefont {Schlag}},\ and\ \bibinfo {author} {\bibfnamefont {M.}~\bibnamefont {Wilhelm}},\ }\bibfield  {title} {\bibinfo {title} {Osmotic engine: Translating osmotic pressure into macroscopic mechanical force via poly(acrylic acid) based hydrogels},\ }\href {https://doi.org/https://doi.org/10.1002/advs.201700112} {\bibfield  {journal} {\bibinfo  {journal} {Advanced Science}\ }\textbf {\bibinfo {volume} {4}},\ \bibinfo {pages} {1700112} (\bibinfo {year} {2017})}\BibitemShut {NoStop}%
\bibitem [{\citenamefont {Choudhary}\ and\ \citenamefont {Raghavan}(2022)}]{Choudhary2022_2}%
  \BibitemOpen
  \bibfield  {author} {\bibinfo {author} {\bibfnamefont {H.}~\bibnamefont {Choudhary}}\ and\ \bibinfo {author} {\bibfnamefont {S.~R.}\ \bibnamefont {Raghavan}},\ }\bibfield  {title} {\bibinfo {title} {Superfast-expanding porous hydrogels: Pushing new frontiers in converting chemical potential into useful mechanical work},\ }\href {https://doi.org/10.1021/acsami.2c00645} {\bibfield  {journal} {\bibinfo  {journal} {ACS Applied Materials \& Interfaces}\ }\textbf {\bibinfo {volume} {14}},\ \bibinfo {pages} {13733} (\bibinfo {year} {2022})},\ \bibinfo {note} {pMID: 35261243}\BibitemShut {NoStop}%
\bibitem [{\citenamefont {Knight}\ and\ \citenamefont {Voth}(2012)}]{Kinght2012_acid1}%
  \BibitemOpen
  \bibfield  {author} {\bibinfo {author} {\bibfnamefont {C.}~\bibnamefont {Knight}}\ and\ \bibinfo {author} {\bibfnamefont {G.~A.}\ \bibnamefont {Voth}},\ }\bibfield  {title} {\bibinfo {title} {The curious case of the hydrated proton},\ }\href {https://doi.org/10.1021/ar200140h} {\bibfield  {journal} {\bibinfo  {journal} {Accounts of Chemical Research}\ }\textbf {\bibinfo {volume} {45}},\ \bibinfo {pages} {101} (\bibinfo {year} {2012})},\ \bibinfo {note} {pMID: 21859071}\BibitemShut {NoStop}%
\bibitem [{\citenamefont {Agmon}(1995)}]{Agmon1995_acid2}%
  \BibitemOpen
  \bibfield  {author} {\bibinfo {author} {\bibfnamefont {N.}~\bibnamefont {Agmon}},\ }\bibfield  {title} {\bibinfo {title} {The grotthuss mechanism},\ }\href {https://doi.org/https://doi.org/10.1016/0009-2614(95)00905-J} {\bibfield  {journal} {\bibinfo  {journal} {Chemical Physics Letters}\ }\textbf {\bibinfo {volume} {244}},\ \bibinfo {pages} {456} (\bibinfo {year} {1995})}\BibitemShut {NoStop}%
\bibitem [{\citenamefont {Wu}\ \emph {et~al.}(1990)\citenamefont {Wu}, \citenamefont {Awakura}, \citenamefont {Ando},\ and\ \citenamefont {Majima}}]{Wu1990Copper}%
  \BibitemOpen
  \bibfield  {author} {\bibinfo {author} {\bibfnamefont {Z.~C.}\ \bibnamefont {Wu}}, \bibinfo {author} {\bibfnamefont {Y.}~\bibnamefont {Awakura}}, \bibinfo {author} {\bibfnamefont {S.}~\bibnamefont {Ando}},\ and\ \bibinfo {author} {\bibfnamefont {H.}~\bibnamefont {Majima}},\ }\bibfield  {title} {\bibinfo {title} {Determination of the diffusion coefficients of cucl2, fecl3, cuso4, and fe2(so4)3 in aqueous solutions},\ }\href@noop {} {\bibfield  {journal} {\bibinfo  {journal} {Materials Transactions Jim}\ }\textbf {\bibinfo {volume} {31}},\ \bibinfo {pages} {1065} (\bibinfo {year} {1990})}\BibitemShut {NoStop}%
\bibitem [{\citenamefont {Marbach}\ and\ \citenamefont {Bocquet}(2019)}]{Marbach2019_2}%
  \BibitemOpen
  \bibfield  {author} {\bibinfo {author} {\bibfnamefont {S.}~\bibnamefont {Marbach}}\ and\ \bibinfo {author} {\bibfnamefont {L.}~\bibnamefont {Bocquet}},\ }\bibfield  {title} {\bibinfo {title} {Osmosis{,} from molecular insights to large-scale applications},\ }\href {https://doi.org/10.1039/C8CS00420J} {\bibfield  {journal} {\bibinfo  {journal} {Chem. Soc. Rev.}\ }\textbf {\bibinfo {volume} {48}},\ \bibinfo {pages} {3102} (\bibinfo {year} {2019})}\BibitemShut {NoStop}%
\bibitem [{\citenamefont {Baker}\ \emph {et~al.}(2010)\citenamefont {Baker}, \citenamefont {Murff},\ and\ \citenamefont {Milam}}]{Baker2010}%
  \BibitemOpen
  \bibfield  {author} {\bibinfo {author} {\bibfnamefont {B.~A.}\ \bibnamefont {Baker}}, \bibinfo {author} {\bibfnamefont {R.~L.}\ \bibnamefont {Murff}},\ and\ \bibinfo {author} {\bibfnamefont {V.~T.}\ \bibnamefont {Milam}},\ }\bibfield  {title} {\bibinfo {title} {Tailoring the mechanical properties of polyacrylamide-based hydrogels},\ }\href {https://doi.org/https://doi.org/10.1016/j.polymer.2010.02.022} {\bibfield  {journal} {\bibinfo  {journal} {Polymer}\ }\textbf {\bibinfo {volume} {51}},\ \bibinfo {pages} {2207} (\bibinfo {year} {2010})}\BibitemShut {NoStop}%
\bibitem [{\citenamefont {Sunyer}\ \emph {et~al.}(2012)\citenamefont {Sunyer}, \citenamefont {Jin}, \citenamefont {Nossal},\ and\ \citenamefont {Sackett}}]{Sunyer2012}%
  \BibitemOpen
  \bibfield  {author} {\bibinfo {author} {\bibfnamefont {R.}~\bibnamefont {Sunyer}}, \bibinfo {author} {\bibfnamefont {A.~J.}\ \bibnamefont {Jin}}, \bibinfo {author} {\bibfnamefont {R.}~\bibnamefont {Nossal}},\ and\ \bibinfo {author} {\bibfnamefont {D.~L.}\ \bibnamefont {Sackett}},\ }\bibfield  {title} {\bibinfo {title} {Fabrication of hydrogels with steep stiffness gradients for studying cell mechanical response},\ }\href {https://doi.org/10.1371/journal.pone.0046107} {\bibfield  {journal} {\bibinfo  {journal} {PLOS ONE}\ }\textbf {\bibinfo {volume} {7}},\ \bibinfo {pages} {1} (\bibinfo {year} {2012})}\BibitemShut {NoStop}%
\bibitem [{\citenamefont {Denisin}\ and\ \citenamefont {Pruitt}(2016)}]{Denisin2016}%
  \BibitemOpen
  \bibfield  {author} {\bibinfo {author} {\bibfnamefont {A.~K.}\ \bibnamefont {Denisin}}\ and\ \bibinfo {author} {\bibfnamefont {B.~L.}\ \bibnamefont {Pruitt}},\ }\bibfield  {title} {\bibinfo {title} {Tuning the range of polyacrylamide gel stiffness for mechanobiology applications},\ }\href {https://doi.org/10.1021/acsami.5b09344} {\bibfield  {journal} {\bibinfo  {journal} {ACS Applied Materials \& Interfaces}\ }\textbf {\bibinfo {volume} {8}},\ \bibinfo {pages} {21893} (\bibinfo {year} {2016})},\ \bibinfo {note} {pMID: 26816386},\ \Eprint {https://arxiv.org/abs/https://doi.org/10.1021/acsami.5b09344} {https://doi.org/10.1021/acsami.5b09344} \BibitemShut {NoStop}%
\bibitem [{\citenamefont {Hossain}\ \emph {et~al.}(2020)\citenamefont {Hossain}, \citenamefont {Roy}, \citenamefont {Sarkar}, \citenamefont {Roy}, \citenamefont {Howlader},\ and\ \citenamefont {Firoz}}]{Hossain2020}%
  \BibitemOpen
  \bibfield  {author} {\bibinfo {author} {\bibfnamefont {M.~A.}\ \bibnamefont {Hossain}}, \bibinfo {author} {\bibfnamefont {C.~K.}\ \bibnamefont {Roy}}, \bibinfo {author} {\bibfnamefont {S.~D.}\ \bibnamefont {Sarkar}}, \bibinfo {author} {\bibfnamefont {H.}~\bibnamefont {Roy}}, \bibinfo {author} {\bibfnamefont {A.~H.}\ \bibnamefont {Howlader}},\ and\ \bibinfo {author} {\bibfnamefont {S.~H.}\ \bibnamefont {Firoz}},\ }\bibfield  {title} {\bibinfo {title} {Improvement of the strength of poly(acrylic acid) hydrogels by the incorporation of functionally modified nanocrystalline cellulose},\ }\href {https://doi.org/10.1039/D0MA00478B} {\bibfield  {journal} {\bibinfo  {journal} {Mater. Adv.}\ }\textbf {\bibinfo {volume} {1}},\ \bibinfo {pages} {2107} (\bibinfo {year} {2020})}\BibitemShut {NoStop}%
\end{thebibliography}
\end{document}